\documentclass[sigconf, 9pt]{acmart}

\setcopyright{none}
\renewcommand\footnotetextcopyrightpermission[1]{}
\makeatletter
\def\@copyrightspace{\relax}
\makeatother

\usepackage{microtype}
\PassOptionsToPackage{warn}{textcomp}
\usepackage{textcomp}
\usepackage{times}

\usepackage{tabu}
\usepackage{booktabs}                  
\usepackage{hyperref}
\usepackage{url}

\usepackage{siunitx}
\usepackage{amsmath,amsfonts}
\usepackage{mathtools, cuted}
\usepackage{algorithmic}
\usepackage[normalem]{ulem}
\usepackage{graphicx}
\usepackage[labelfont=normalfont,textfont=normalfont]{caption}
\usepackage{subcaption}
\usepackage{xcolor}
\usepackage{float}
\usepackage{tabularx}
\usepackage{array}
\usepackage{balance}
\usepackage{multirow}
\newcolumntype{P}[1]{>{\centering\arraybackslash}p{#1}}
\newcolumntype{M}[1]{>{\centering\arraybackslash}m{#1}}

\newenvironment{myitemize}{\begin{list}{$\bullet$}{\renewcommand{\leftmargin}{0.15in}}}{\end{list}}

\begin{document}

\title{Here to Stay: Measuring Hologram Stability in Markerless Smartphone Augmented Reality}

\author{Tim Scargill}
\affiliation{%
  \institution{Duke University}
  \streetaddress{}
  \city{}
  \country{}}
\email{ts352@duke.edu}

\author{Jiasi Chen}
\affiliation{%
  \institution{University of California, Riverside}
  \streetaddress{}
  \city{}
  \country{}}
\email{jiasi@cs.ucr.edu}

\author{Maria Gorlatova}
\affiliation{%
  \institution{Duke University}
  \streetaddress{}
  \city{}
  \country{}}
\email{maria.gorlatova@duke.edu}

\begin{abstract}
Markerless augmented reality (AR) has the potential to provide engaging experiences and improve outcomes across a wide variety of industries; the overlaying of virtual content, or \emph{holograms}, onto a view of the real world without the need for predefined markers provides great convenience and flexibility. However, unwanted hologram movement frequently occurs in markerless smartphone AR due to challenging visual conditions or device movement, and resulting error in device pose tracking. We develop a method for measuring hologram positional errors on commercial smartphone markerless AR platforms, implement it as an open-source AR app, HoloMeasure, and use the app to conduct systematic quantitative characterizations of hologram stability across 6 different user actions, 3 different smartphone models, and over 200 different environments. 
Our study demonstrates significant levels of spatial instability in holograms in all but the simplest settings, and underscores the need for further enhancements to pose tracking algorithms for smartphone-based markerless AR. 
\end{abstract}





\maketitle
\pagestyle{plain}

\section{Introduction}
\label{sec:Introduction}
Augmented reality (AR) is a revolutionary 
technology with the potential to transform the way we interact with computers across a wide range of industries, from healthcare and design to education and entertainment. The overlaying of virtual content or \emph{holograms} onto a view of the real world provides engaging experiences which can inform, instruct, and entertain. However the quality of these experiences, and the likelihood that someone uses AR again in the future after trying it out once or twice, is heavily influenced by a critical factor: that users, especially new ones, expect holograms to appear and behave like real objects. Along with accurate light rendering \cite{rohmer2017natural,prakash2019gleam} and occlusion effects \cite{hebborn2017occlusion,holynski2018fast}, one of the most noticeable aspects of this is \emph{spatial stability}: a hologram should remain exactly where a user places it, even if in the meantime the user moves around, walks away, pauses the AR app to send a text message, or performs any other action. When the hologram does not, the illusion is immediately broken; users may become disengaged or frustrated, and hologram instability impedes or compromises the task they are performing. If AR truly is “here to stay”, then the holograms that we place within it must be as well.

In the last few years \emph{markerless} AR has become commonplace. Markerless AR enables the positioning of virtual content without predefined markers (easily recognizable textures such fiducial markers, similar to QR codes, or images); natural features within the environment are mapped and tracked instead. The convenience this markerless technique provides has led to its widespread use in recent AR apps, such as Pokemon GO~\cite{PokemonGo}, IKEA Place~\cite{IKEAPlace}, Augment~\cite{Augment}, and Angry Birds AR~\cite{AngryBirds}. The vast majority of modern AR platforms, such as \mbox{ARCore}~\cite{ARCore} and \mbox{ARKit}~\cite{ARKit}, support markerless AR through \emph{\mbox{Visual-Inertial} Simultaneous Localization and Mapping \mbox{(VI-SLAM)}}. VI-SLAM is used to concurrently map the environment and track the pose (position and orientation) of the AR device. If the correct pose is known in relation to the surrounding environment, then holograms can be rendered in the correct position relative to the real world.

\begin{figure}
\captionsetup[subfigure]{justification=centering}
\begin{subfigure}{.155\textwidth}
  \includegraphics[width=0.95\linewidth]{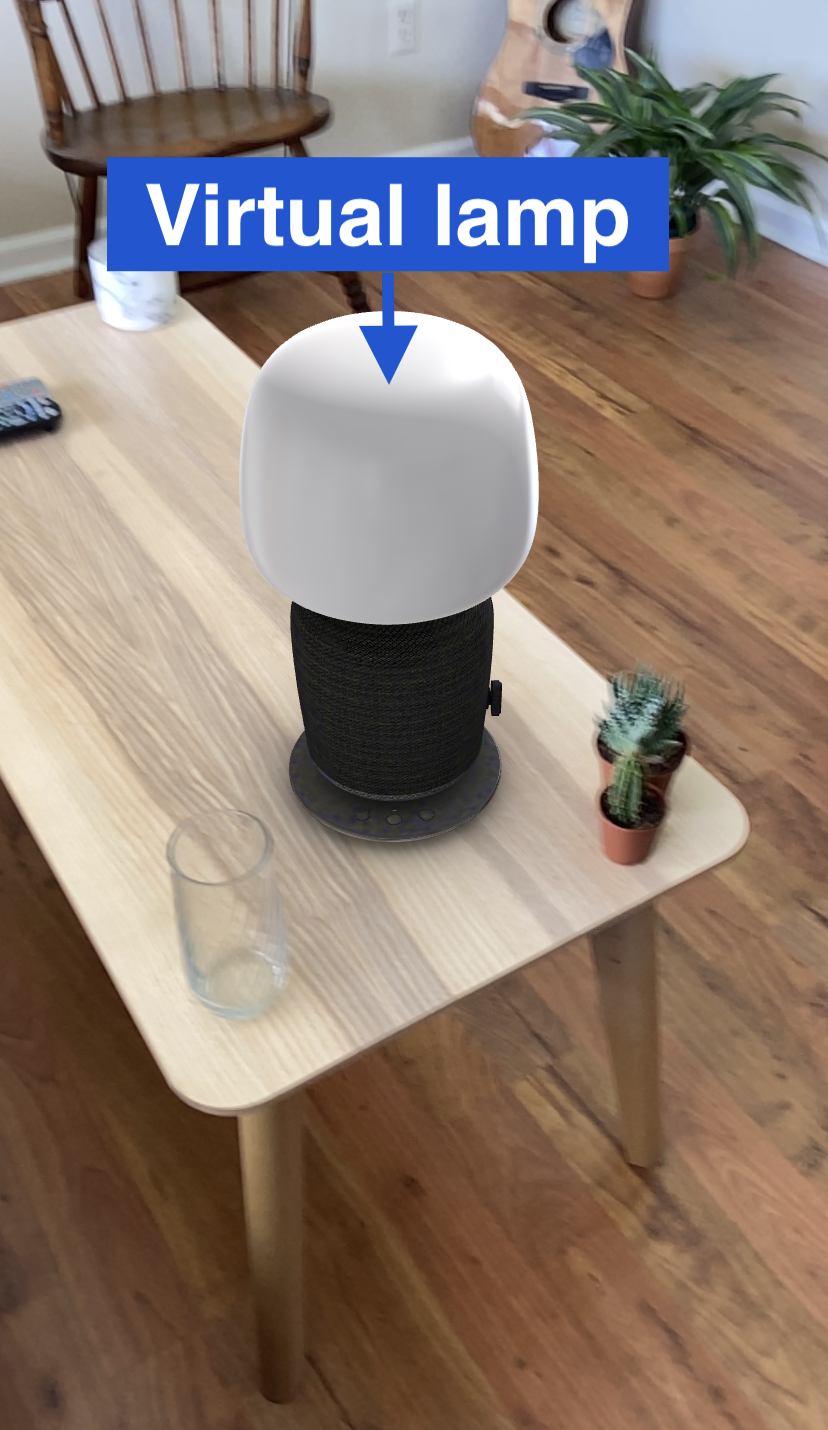}
  \vspace{-0.03in}
  \caption{Time 0}
  \label{fig:Drift0}
\end{subfigure}
\begin{subfigure}{.155\textwidth}
  \includegraphics[width=0.95\linewidth]{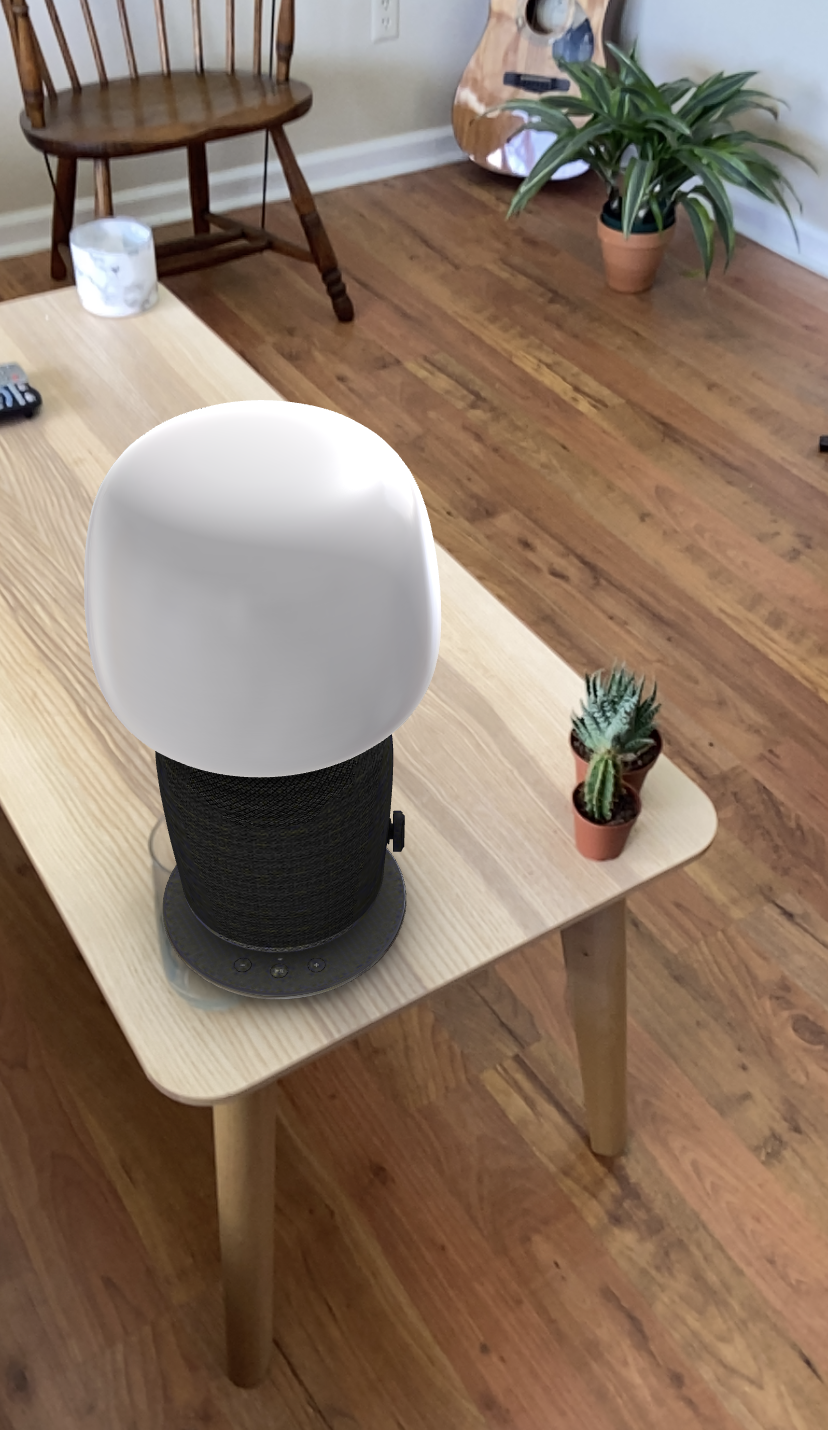}
  \vspace{-0.03in}
  \caption{Time 1}
  \label{fig:Drift1}
\end{subfigure}
\begin{subfigure}{.155\textwidth}
  \includegraphics[width=0.95\linewidth]{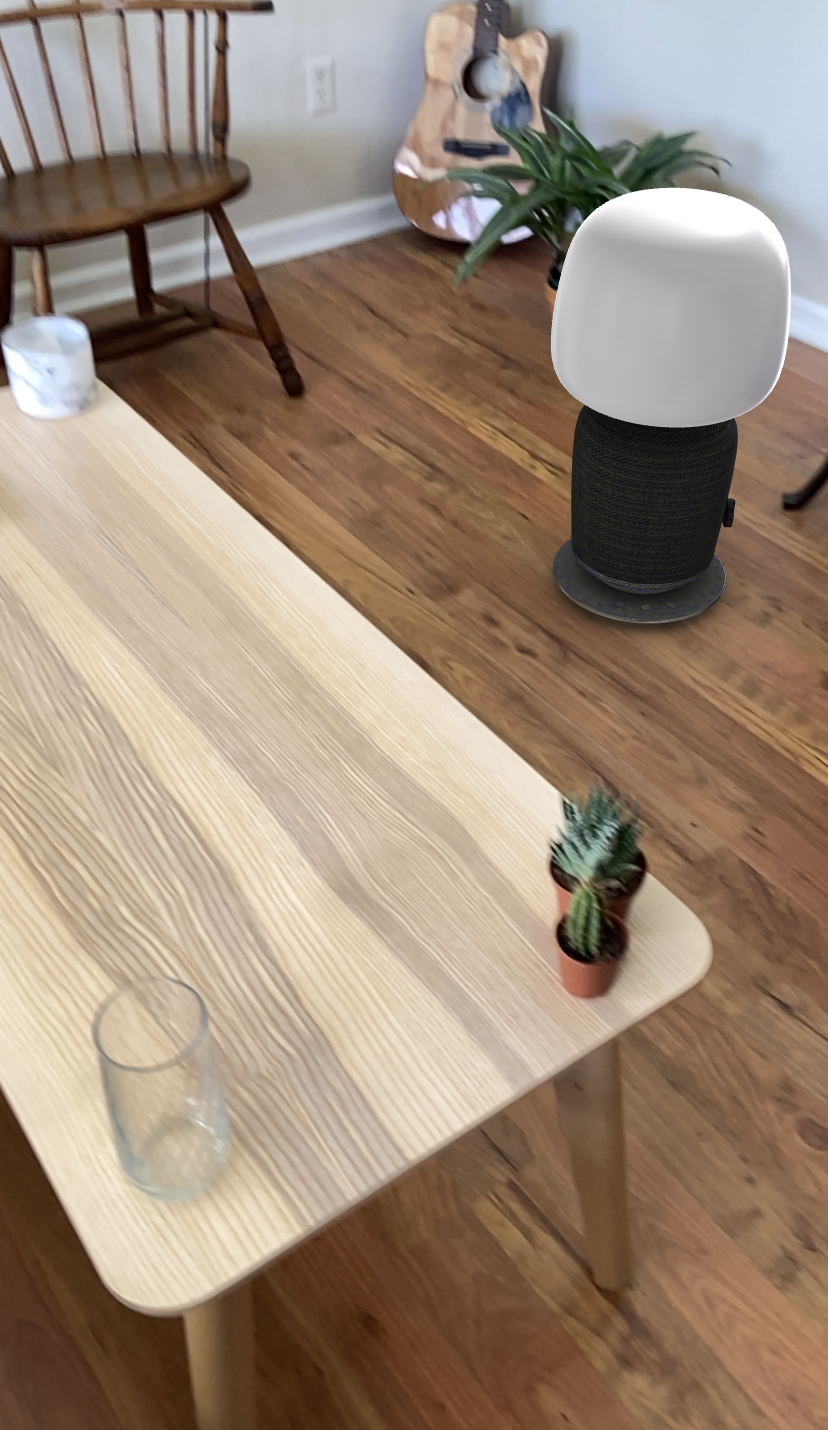}
  \vspace{-0.03in}
  \caption{Time 2}
  \label{fig:Drift2}
\end{subfigure}
\vspace{-0.12in}
\caption{Hologram drift in markerless \mbox{VI-SLAM-based AR} (ARKit, iPhone 11, iOS 14.4, IKEA Place app \cite{IKEAPlace}). A virtual lamp drifts from its original position (a), colliding with a real-world glass on the table (b), and moving off the table entirely (c).}
\label{fig:DriftExamples}
\end{figure}

However, errors still frequently occur in the pose estimation process because of challenging visual conditions or device movement. This results in a hologram being rendered in a different position from where it was originally placed, a problem often termed \textit{drift} (i.e., the Euclidean distance between the original and resulting hologram positions). Examples of drift in markerless AR are shown in Figure~\ref{fig:DriftExamples}. At `Time 0' (Figure~\ref{fig:Drift0}) the user places the virtual black and white lamp on the table; they then walk away from the area (perhaps to bring another person to come and look at the lamp) before returning to view it at `Time 1'. The virtual lamp has drifted out of position and collided with a real-world glass (Figure~\ref{fig:Drift1}). After walking away and returning again, the drift is even more pronounced at `Time 2', and the lamp has now moved off the table (Figure~\ref{fig:Drift2}). These positional errors can have a severe negative impact on a user's experience, especially when the hologram violates real-world physics. 

Hologram stability errors are a much more noticeable issue on smartphones than modern AR headsets such as the Microsoft Holo-Lens \cite{HoloLens} or Magic Leap One~\cite{MagicLeapOne}. For example, while we will show in our experiments that drift magnitude is frequently greater than 5cm on smartphones, a small scale study on the Microsoft Holo-Lens~1~\cite{vassallo2017hologram} found that the drift values after performing 4 different device movements were between 0.5cm and 0.6cm. This is because the form factor of a headset allows for both the inclusion of specialized environment sensors, and their arrangement such that they face in multiple directions, improving tracking quality. 
Because the most common type of AR experience is currently `single-session', i.e. virtual content does not persist once an application is closed, we test this type of scenario here. Finally, for feasibility we also limit the scope of this work to static indoor environments; future work should examine the challenges associated with other types of environments.  

Our goal is to measure spatial stability of the holograms in markerless smartphone AR for different device movements, in a wide variety of indoor environments. We define two metrics for hologram instability. We use \textit{drift} to refer to the measurable displacement when a hologram is rendered for an extended period in a different location. Sometimes the AR system may also lose tracking temporarily, and during the period in which the device attempts to relocalize, unwanted hologram movement frequently occurs. We term this period of instability \textit{relocalization time}. While relocalization time can be captured directly in ARCore and ARKit, current methods for measuring drift either require optical tracking systems, the setup of which is infeasible for large numbers of different environments, or use fiducial markers to measure the position of the hologram \cite{ran2019sharear}, which artificially improves the environment under test by introducing a recognizable texture. We therefore require a new method to measure hologram drift.  

To this end we develop an open-source, cross-platform app called \mbox{HoloMeasure}\footnote{\label{note1}Our AR session measurement app HoloMeasure is available on GitHub at \url{https://github.com/SceneIt-Source/HoloMeasure}}, that captures environment characteristics, inertial data, device tracking data, hologram drift and relocalization time on ARKit and ARCore, and evaluate its accuracy in terms of drift measurement (Section~\ref{sec:ARPerformanceMeasurementApp}). We examine a set of 6 common actions a user may perform in smartphone AR for the scenario of viewing a stationary hologram (Section~\ref{subsec:DeviceMovementinSmartphoneAR}), and conduct experiments to compare performance across multiple platforms and devices \mbox{(Section~\ref{subsec:CrossPlatformHologramStabilityExperiment})}. We then measure hologram stability in wide variety of visual environments on an iPhone 11 running iOS 14.4 and ARKit 4 (Section~\ref{subsec:ARVisualEnvironmentsDataset}), and additionally analyze performance for those environments which are consistent with ARKit guidelines \cite{ARKitGuidelines} (Section~\ref{subsec:RecommendedARKitVisualEnvironments}). We then discuss 
the implications of our findings (Section~\ref{sec:Discussion}) and present our conclusions (Section~\ref{sec:Conclusion}). 
Our key contributions can be summarized as follows:
\begin{myitemize}
  \item Using Unity's AR Foundation \cite{ARFoundation} we develop an open-source AR app, HoloMeasure, available on GitHub\footnotemark[1], that can be used to measure environment properties, device movement, and hologram stability on ARKit and ARCore. In 95\% of cases, the app can achieve a drift measurement accuracy of 0.4cm.  
  
  \item We conduct cross-platform experiments on ARKit and ARCore with 3 smartphones, for 6 actions in 20 environments, revealing measurable and highly variable drift across all platforms, devices and actions. ARKit achieves lower hologram drift on 4 of 6 actions, but ARCore performs significantly better when a user walks away and returns to a hologram (mean drift of 4.8cm on ARCore and 25.1cm on ARKit).
  
  
  \item We perform a hologram stability study in a wide variety of visual environments on ARKit, in over 200 environments for each of our 6 actions. We show that hologram stability is highly dependent on both the action a user performs between hologram views (with mean drift ranging from 1.2cm to 43.2cm for our set of 6 actions) and the visual characteristics of the entire environment (e.g., lighting, texture), both where the hologram is placed and in surrounding areas. 
  
 
\end{myitemize}


To the best of our knowledge, this work is the first to perform a quantitative side-by-side comparison of hologram stability on different software platforms and smartphone models, and the first to measure hologram stability in a large number of different environments, incorporating a wide range of visual conditions.

\section{Related Work}
\label{sec:RelatedWork}
\textbf{AR platform tracking state:} ARCore and ARKit both provide indicators of current tracking quality \cite{ARCoreTrackingState, ARKitTrackingState}, though granularity is limited and neither include the magnitude of hologram instability likely to occur. ARKit indicates whether tracking is `not available', `limited' (results are questionable), or `normal', while ARCore describes the tracking state as `stopped', `paused' (which may indicate that device tracking has been lost), or `tracking'. The method by which these tracking quality states are derived is proprietary. ARKit and ARCore provide recommendations for visual environment characteristics in their documentation \cite{ARKitGuidelines, ARCoreGuidelines}, but both are in the context of enabling accurate plane detection rather than tracking state or hologram stability. There is also an ARCore tool for checking the quality of reference images for image recognition \cite{arcoreimg} (again the assessment criteria are proprietary), but this is a type of marker-based AR in which device pose is detected from 2D image features, rather than markerless AR based on natural features. \emph{To the best of our knowledge, no data on virtual content positional accuracy achieved on these platforms for markerless AR has been made public}.


\textbf{Virtual content positional errors:} While virtual content positional errors are a known problem in AR \cite{ran2019sharear, vassallo2017hologram}, current methods of assessing them on commercial platforms have significant limitations. One study \cite{blom2018impact} uses a subjective assessment of whether holograms remain stationary in different lighting conditions, but does not quantify errors. Another work \cite{vassallo2017hologram} uses an optically-tracked stylus to measure the positions of each corner of a hologram before and after movement to calculate drift. While the tracking system provides sub-millimeter accuracy \cite{PolarisVega}, the human operator also introduces further measurement error. Even more importantly, the logistics involved in the setup and calibration of the optical tracking system make studying a wide variety of environments infeasible. In \cite{ran2019sharear}, the authors propose a technique that uses fiducial markers as points of reference from which errors can be calculated. Single fiducial markers can provide pose accuracy up to approximately 0.5cm at a distance of 0--2m \cite{kalaitzakis2021fiducial}, but these markers introduce recognizable textures into the environment, which improve tracking accuracy and compromise our ability to assess performance in challenging visual conditions. In a recent demo \cite{scargill2021will}, we used the hologram drift measurement method described in this report (Section~\ref{sec:ARPerformanceMeasurementApp}) to capture virtual content position error and visual conditions in 141 environments, and developed a visual environment binary classifier based on drift magnitude.

\textbf{Pose tracking accuracy:} Another option is to assess virtual content positional errors indirectly by measuring the pose tracking accuracy of an AR system; this requires external ground truth measurements. Ground truth in \cite{cortes2018advio} is obtained through a combination of a purely inertial odometry system plus known manual fixation points. This facilitates measurements in a variety of environments, but only to dm--m level position accuracy, which is not sufficient for assessing virtual content positional errors. Ground truth position can be obtained to sub-millimeter and orientation to sub-degree level accuracy by using a motion capture system such as OptiTrack \cite{OptiTrack} or Vicon \cite{Vicon} (although accuracy has been shown to be lower at higher velocities, along the axis of movement \cite{furtado2019comparative}). This is the standard method for measuring the performance of VI-SLAM systems \cite{burri2016euroc, schubert2018tum, jinyu2019survey, kasper2019benchmark, zuniga2020vi}; there has been much valuable work in the creation of benchmark datasets, against which new algorithms can be evaluated. The current best-performing open-source VI-SLAM algorithm for the most well-known benchmarks, the EuRoC drone dataset \cite{burri2016euroc} and the TUM VI handheld dataset \cite{schubert2018tum}, is ORB-SLAM3~\cite{campos2021orb}, which achieves an average position estimate accuracy of 3.5cm for EuRoC and 0.9cm for TUM~VI. We cannot use these datasets to evaluate the VI-SLAM algorithms in commercial AR platforms because those algorithms are closed-source, though we can use the same method for capturing ground truth, by attaching a rigid set of retroreflective markers to the AR device, as in \cite{feigl2020localization}, which examined performance in a large-scale industrial environment. However, any use of a motion capture system requires the entire area the device moves through to be visible from multiple specialized cameras, and again the associated setup and calibration logistics means that studying a large number of different environments is not practical (EuRoC only contains 3 visual environments with accurate ground truth, and TUM~VI only 1).



\section{AR Session Measurement App}
\label{sec:ARPerformanceMeasurementApp}
In this section we develop an open-source smartphone app called \mbox{HoloMeasure}, that can be used to measure environment properties, device motion, tracking status, and hologram stability during an ARKit or ARCore session. Our app is available on GitHub at \mbox{\url{https://github.com/SceneIt-Source/HoloMeasure}}.

\subsection{Motivation}
Hologram stability errors arise from errors in device pose tracking, accomplished using a VI-SLAM algorithm. Traditional VI-SLAM performance evaluations \cite{burri2016euroc, schubert2018tum, jinyu2019survey, kasper2019benchmark, zuniga2020vi} are informative, but alone they are not ideal for studying hologram stability in mobile~AR. To achieve the accuracy needed, these evaluations require precise ground truth pose data, recorded with a motion capture system such as OptiTrack \cite{OptiTrack} or Vicon~\cite{Vicon}. While these motion capture systems can achieve 0.1cm level accuracy for positional error in optimal conditions~\mbox{\cite{furtado2019comparative, jinyu2019survey}}, the time and logistics involved in the transport and placement of at least 4 tripod-mounted cameras, connecting those cameras via cables to a nearby computer, and moving the calibration wand around the measurement area, for each new setting, mean only a small number of different visual environments is feasible. For a computing platform that is intended to be ubiquitous, for deployment wherever the user takes their smartphone, this is insufficient. Alternatively, hologram drift can be calculated by measuring the position of virtual content relative to fiducial markers~\cite{ran2019sharear}. This method can provide 0.5cm accuracy~\cite{kalaitzakis2021fiducial}, but modifies the natural environment under test by introducing additional recognizable textures. This is likely to improve pose tracking performance, and removes the ability to study challenging featureless environments. We note, however, that drift can also be calculated directly, from the distance between where the hologram was first placed and where it moved to throughout the course of the session. We leverage this for our drift measurement methodology.


\subsection{Measurement App Design}
\label{subsec:MeasurementAppDesign}
With this in mind, we have developed an open-source, cross-platform app, HoloMeasure, to measure hologram drift on commercial AR platforms. Developed using Unity's AR Foundation framework \cite{ARFoundation}, HoloMeasure comes in the form of a Unity project which can be built and run on iOS or Android devices. Consistent with the latest versions of ARKit and ARCore, holograms can be placed wherever a plane is detected in the environment, but the design will also support the case of holograms attached to a world mesh. In the background the app captures data on the visual environment (through camera images sampled during mapping and the point cloud generated during mapping), inertial data from the accelerometer and gyroscope, device position (relative to the start position), and the tracking status of the AR platform (including relocalization time). Drift is calculated through the user's placement of holograms, as described below.



\subsection{Measurement App Use}
\label{subsec:MeasurementAppUse}
\begin{figure}
\centering
\captionsetup[subfigure]{justification=centering}
\begin{subfigure}{.155\textwidth}
  \includegraphics[width=0.95\linewidth]{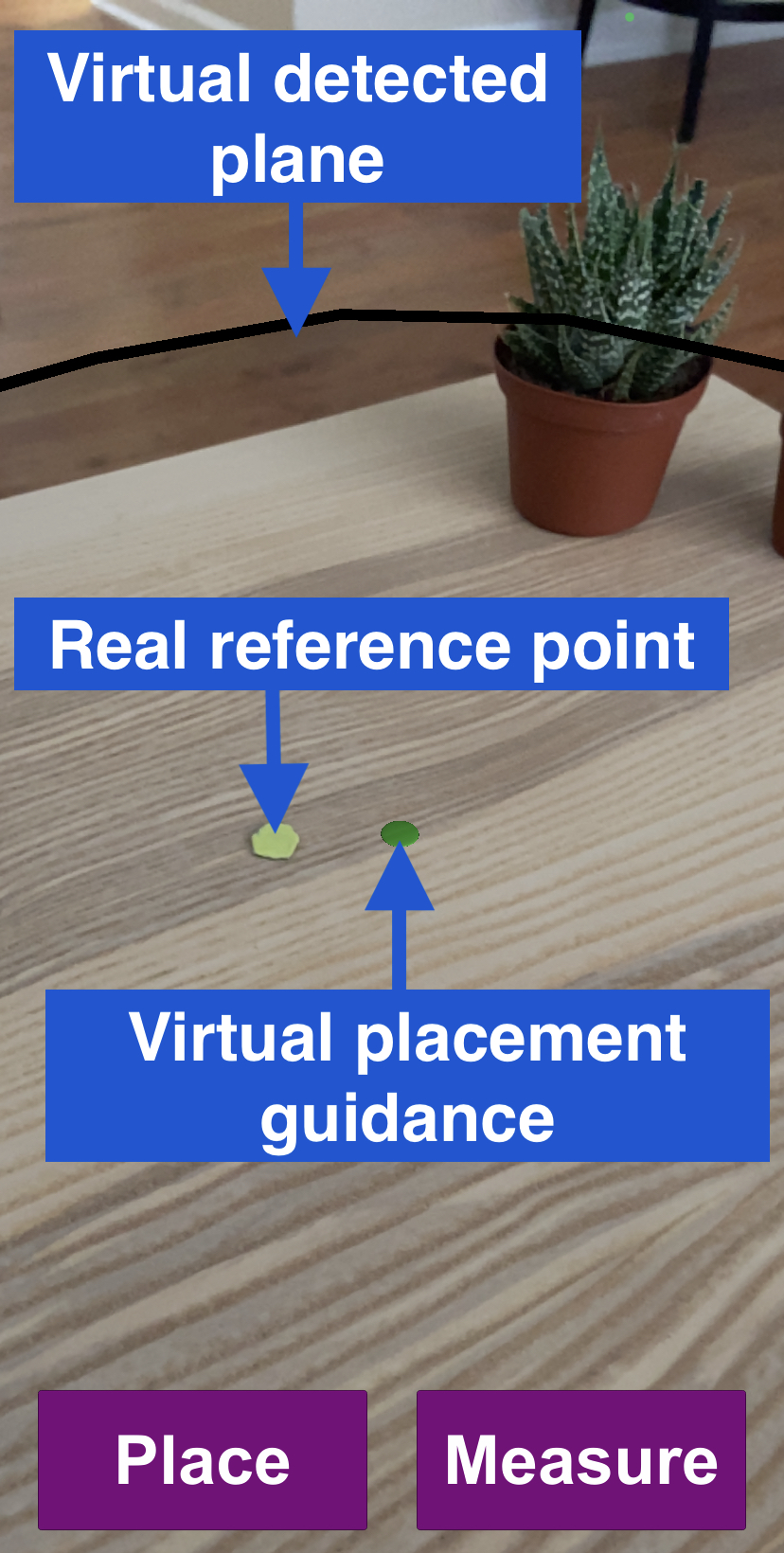}
  \vspace{-0.08in}
  \caption{App interface}
  \label{fig:Tool1}
\end{subfigure}
\begin{subfigure}{.155\textwidth}
  \includegraphics[width=0.95\linewidth]{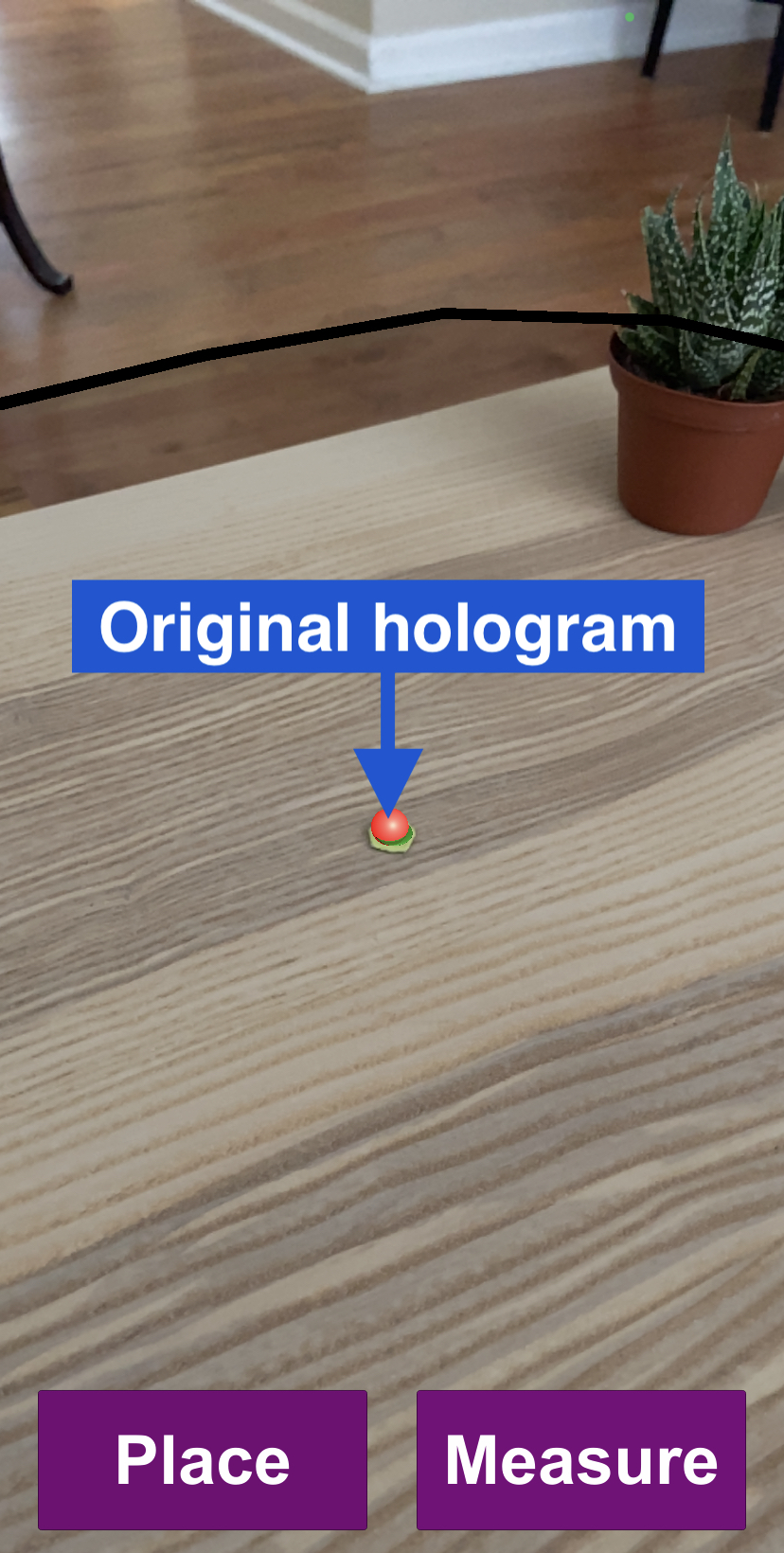}
  \vspace{-0.08in}
  \caption{Placed hologram}
  \label{fig:Tool2}
\end{subfigure}
\begin{subfigure}{.155\textwidth}
  \includegraphics[width=0.95\linewidth]{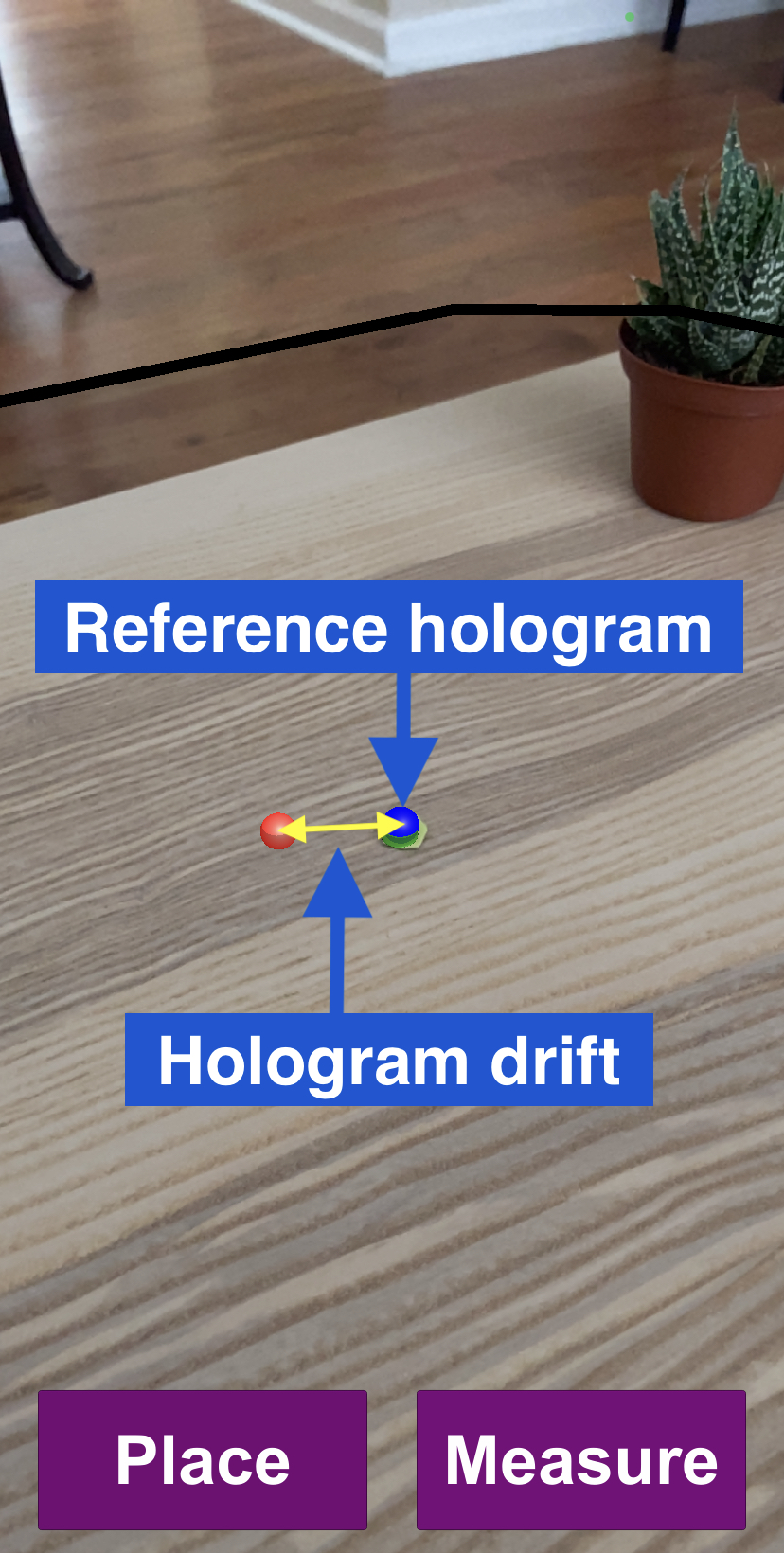}
  \vspace{-0.08in}
  \caption{Placed reference}
  \label{fig:Tool3}
\end{subfigure}
\vspace{-0.08in}
\caption{Our AR session measurement app, HoloMeasure. The user places a real reference point (e.g., a small sticker or piece of paper) on a real-world surface (a) and places the original hologram on it using the app (b). The user performs an action, such as moving to view the hologram from a different angle, then measures the drift of the original hologram by placing a reference hologram on the real reference point (c).}
\label{fig:TestingTool}
\end{figure}

With visual, inertial and tracking data captured in the background, interaction with our measurement app is centered around placing holograms at a clearly defined position in the real world, marked by a real reference point such as a sticker or piece of paper (see Figure~\ref{fig:Tool1}). The size of the real reference point and the holograms must make them clearly visible when viewed within the range of holograms placement distances (Euclidean distance from the camera to the hologram) that the user wishes to study. However, the real reference point should not be so large that it covers a significant proportion of the real world textures, modifying the environment unnecessarily. We also find that the diameter of the real reference point should be approximately 25\% larger than the size of the hologram. This is because the virtual placement guidance (a green circle that indicates where a hologram will be placed) and holograms may sometimes be rendered larger than intended due to underestimation of the distance from the camera to the plane by the AR platform. When the virtual content appears larger than and obscures the real reference point, accurate placement is severely impeded, but the extra 25\% in diameter of the real reference point avoids the vast majority of these cases.

To start session measurement the user opens the app and mimics a normal AR user, moving the device to detect a plane while remaining focused on the real reference point. Virtual placement guidance, a green virtual circle, indicates where a hologram will be placed, and the user presses the `Place' button to position a red sphere on the real reference point (Figure~\ref{fig:Tool2}). The user presses the `Measure' button to confirm the hologram placement position. They then perform the movement or actions of their choice before measuring drift. The user places a blue sphere on the reference point (Figure~\ref{fig:Tool3}). If the original red sphere has not drifted away from the real reference point, the two holograms will collide, but if it has drifted, then there will be a displacement between the two. When the user presses `Measure', the Euclidean distance, in world space as computed by the AR platform, between the two spheres is calculated. Based on our observations, we estimate that this manual alignment of holograms with a reference point facilitates drift measurement accuracy to within 0.4cm (see Section~\ref{subsec:MeasurementAppAccuracy} for details). The process can be repeated rapidly in any location where the user can temporarily place a small marker, enabling them to quickly build a larger, more varied dataset of environments and their associated hologram drifts than traditional VI-SLAM benchmarks.

\subsection{Drift Measurement Accuracy Evaluation}
\label{subsec:MeasurementAppAccuracy}
\begin{figure}
\captionsetup[subfigure]{justification=centering}
\begin{subfigure}{.170\textwidth}
  \includegraphics[width=1\linewidth]{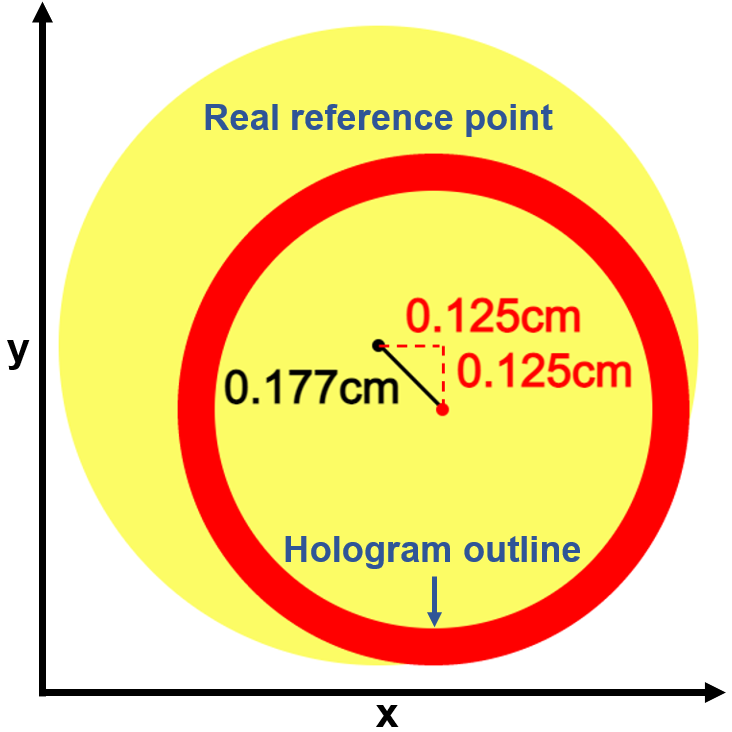}
  \vspace{-0.04in}
  \caption{Theoretical hologram placement accuracy}
  \label{fig:PlacementAccuracyDiagram}
\end{subfigure}
\begin{subfigure}{.280\textwidth}
  \includegraphics[width=1\linewidth]{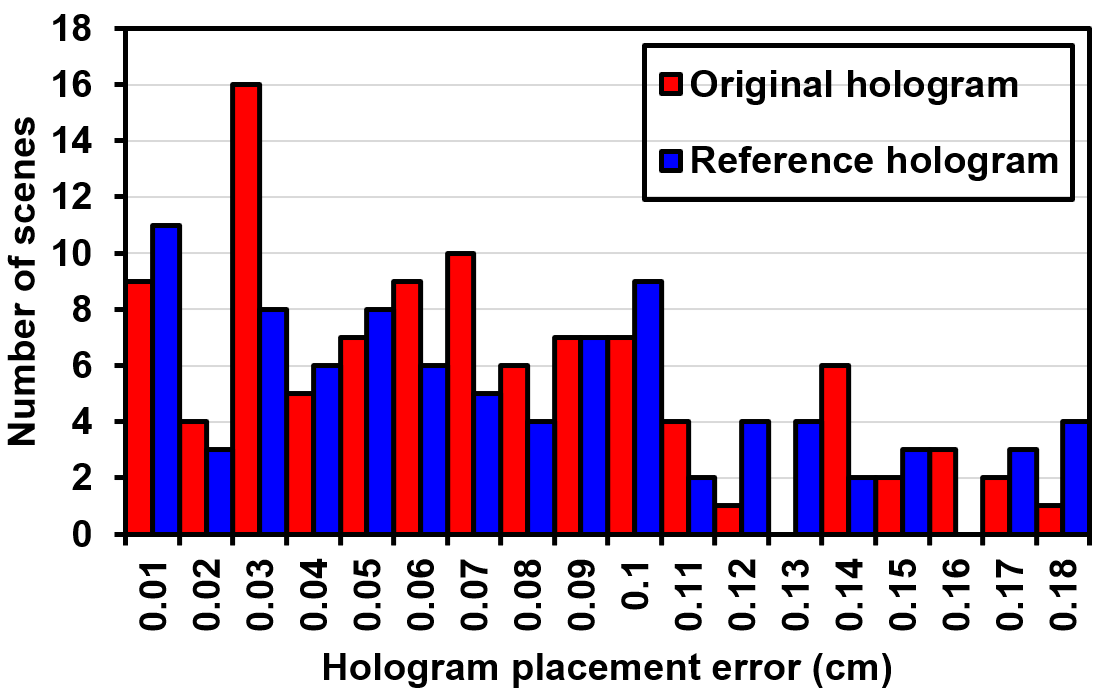}
  \vspace{-0.04in}
  \caption{Histogram of hologram placement errors}
  \label{fig:PlacementAccuracyHistogram}
\end{subfigure}
\vspace{-0.1in}
\caption{Theoretical hologram placement accuracy (0.177cm) when hologram is placed within the bounds of the yellow real reference point (a), and histogram of measured placement errors from 100 scenes (b), both for a real reference point with diameter 1.25cm and holograms of diameter 1cm.}
\label{fig:MeasurementAccuracy}
\end{figure}

While timing-based performance metrics such as relocalization time are captured automatically by monitoring the tracking state of the AR platform, our hologram drift measurement method introduces a source of error because it relies on the manual placement of holograms. The accuracy of our method is defined by the difference in diameter between the real reference point and the holograms, along with how often the user is able to place the holograms within the bounds of the real reference point. The latter is largely defined by an appropriate choice of hologram and real reference point sizes, along with the user's experience with the app and AR in general. In our experiments we wished to study hologram placement distances of 0.2--2.5m, and we found that holograms of 1cm diameter and a real reference point of 1.25cm diameter best suited these conditions. In this case, assuming the user places the hologram within the bounds of the real reference point, then the maximum errors on the x and y axes are 0.125cm, the difference between the radii of the real reference point and the hologram. Therefore the maximum total error for these axes combined is 0.177cm, as illustrated in Figure~\ref{fig:PlacementAccuracyDiagram}. The errors from the original and reference hologram placements could combine, giving a potential total manual placement error of 0.354cm, when the hologram is placed within the bounds of the real reference point.


To substantiate this manual placement error bound in our experiments we used the two screenshots captured by HoloMeasure when the original and the reference holograms placements were finalized. We analyzed 100 scenes, covering a representative range of environments, placement distances and light conditions, giving us 200 total hologram placements. First we recorded the proportion of occasions when the hologram was placed within the bounds of the real reference point, viewed by zooming into the screenshot. For 190 of 200 screenshots, 95\% of placements, the hologram was placed within the real reference point. We also calculated the placement error by manually measuring the number of pixels visible of the real reference point on either side of the placed hologram. If the number of pixels either side of the hologram is equal, then there is `zero' error (this method can only guarantee that the error is less than 1mm, the maximum pixel width in world space for our screenshots), but if the number of pixels is unequal, then we identify a case of manual placement error. In those cases we calculate the number of pixels by which the hologram is off-center, then multiply that by the width in world space of a pixel for that image (calculated by dividing the diameter of the real reference point, 1.25cm, by its diameter in pixels), to obtain an error estimate. A histogram of the manual placement errors in our 100 scenes (excluding those 10 occasions when the hologram was placed outside the real reference point) is shown in Figure~\ref{fig:PlacementAccuracyHistogram}; for all values the measured placement error was less than 0.18cm. This gives a total maximum error of 0.36cm for each scene, and so given the accuracy of our pixel-based error measurement method, we estimate that an experienced user of HoloMeasure can achieve 0.4cm drift measurement accuracy in 95\% of cases. This is comparable to using fiducial markers for the same distance range, 0.5cm \cite{kalaitzakis2021fiducial}, but does not modify the natural environment under test by introducing additional recognizable textures. It is less accurate than using optical tracking systems (0.1cm accuracy \cite{furtado2019comparative, jinyu2019survey}), but our methodology is much faster and more practical for studying a range of different environments, because it does not require the setup and calibration of external cameras.

An additional insight we obtained from this measurement of manual placement error is that the accuracy of hologram placement differed for the original and reference holograms. Only 1 of the 100 original holograms was placed outside the bounds of the real reference point, but 9 out of 100 for the reference holograms. Furthermore the mean placement error for the remaining holograms was 0.07cm for the original holograms, and 0.1cm for the reference holograms. While this difference might be attributed to less care being taken on the second hologram placement, we believe there may be another reason for this. When the original hologram is placed the view of the real reference point is unobstructed, but because in this version of the app we continue to render the original hologram when the reference hologram is placed, the view at that time may sometimes be obstructed, causing this greater manual placement error. Therefore a potential way to improve the accuracy of our method would be to stop rendering the original hologram after its position has been confirmed.

\section{Hologram Stability Experiments}
\label{sec:ARPlatformEvaluations}
In this section we use our AR session measurement app, HoloMeasure (Section~\ref{sec:ARPerformanceMeasurementApp}), to conduct a set of hologram stability experiments on commercial AR platforms. In Section~\ref{subsec:DeviceMovementinSmartphoneAR} we define a set of actions 
users may potentially perform while viewing a hologram; in Section~\ref{subsec:CrossPlatformHologramStabilityExperiment} we compare performance across different platforms and devices, in Section~\ref{subsec:ARVisualEnvironmentsDataset} we measure hologram stability in a wide variety of diverse environments on ARKit, and in Section~\ref{subsec:RecommendedARKitVisualEnvironments} we comment on hologram stability in the subset of 
our ARKit environments that excludes insufficiently lit and insufficiently textured scenes. 

\subsection{Device Movement in Smartphone AR}
\label{subsec:DeviceMovementinSmartphoneAR}

We consider 6 actions a user may perform after placing a stationary hologram:
\begin{myitemize} 
\item \textit{Focused move}: the user moves to a different viewing angle (to inspect another side of the hologram) while keeping the camera focused on the hologram.

\item \textit{Unfocus}: the user lets the hand holding the device hang at their side (as if distracted), then raises the device to focus on the hologram.

\item \textit{Unfocused move}: the user lets the hand holding the device hang at their side, moves to a different viewing angle, then raises the device to focus on the hologram.

\item \textit{Walk away}: the user lets the hand holding the device hang at their side, walks away (as if to e.g., perform another action), returns and raises the device to focus on the hologram.

\item \textit{Pause}: the user pauses the AR app (as if to e.g., send a text), then resumes the session while focused on the hologram position.

\item \textit{Place down}: the user places the device down on a nearby surface temporarily so that the rear camera is covered (as if to e.g., pick up another object), then raises the device to focus on the hologram.
\end{myitemize}

Our goal is to examine the extent to which these actions cause hologram instability on different devices, and in different environments. As well as measuring the drift observed after each action, the \textit{Pause} and \textit{Place down} actions incur a temporary loss of tracking, so we can also examine \emph{relocalization time} in these cases. 

\begin{figure}
\captionsetup[subfigure]{justification=centering}
\begin{subfigure}{.230\textwidth}
  \includegraphics[width=1\linewidth]{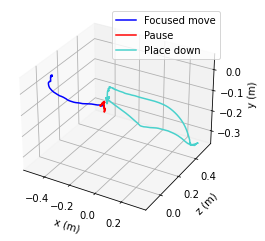}
  \vspace{-0.04in}
  \caption{}
  \label{fig:Movements1}
\end{subfigure}
\begin{subfigure}{.230\textwidth}
  \includegraphics[width=1\linewidth]{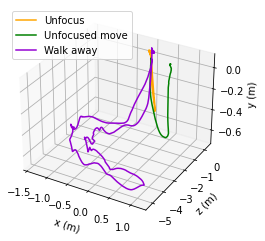}
  \vspace{-0.04in}
  \caption{}
  \label{fig:Movements2}
\end{subfigure}
\vspace{-0.1in}
\caption{Representative device trajectories for 6 common actions that users may perform after hologram placement: \textit{Focused move}, \textit{Pause}, and \textit{Place down} (a), \textit{Unfocus}, \textit{Unfocused move}, and \textit{Walk away} (b).}
\label{fig:DeviceMovements}
\end{figure}

For all subsequent experiments, the change in viewing angle for \textit{Focused move} and \textit{Unfocused move} was set to approximately $45^{\circ}$ (on a horizontal plane), the distance walked away in \textit{Walk away} to 10 steps (approximately 7m), and the \textit{Pause} and \textit{Place down} interruption duration to approximately 5s. Representative trajectories for each of these actions, captured in our experiments, are shown in Figure~\ref{fig:DeviceMovements}.

\subsection{Cross-Platform Experiments}
\label{subsec:CrossPlatformHologramStabilityExperiment}
To examine how robust different AR platforms and devices are to these 6 actions, we carried out a side-by-side comparison with 3 devices, an iPhone~11 (iOS 14.4 and ARKit 4), a Samsung Galaxy Note 10+ and a Nokia 7.1 (both Android 11 and ARCore 1.23). Relevant hardware and software specifications are shown in Table~\ref{tab:DeviceComparison}. We performed each action on each device in 20 environments, captured in 6 rooms in 2 buildings, for a total of 360 scenes.\footnote{We adopt the terminology \textit{environment} to refer to the space in which an AR session takes place, and \textit{scene} to refer to a single AR session.} 16 of these 20 environments are consistent with the ARKit and ARCore guidelines \cite{ARKitGuidelines,ARCoreGuidelines} for well-lit and textured environments, and none were extremely challenging (e.g., dark conditions, or blank surfaces). To ensure a fair comparison, in each environment we performed actions consecutively from exactly the same positions, under fixed artificial lighting, and no other applications were run concurrently with our measurement app. \emph{To the best of our knowledge this study is the first published quantitative, side-by-side comparison of hologram stability on different software platforms and smartphone models.}

\begin{table}[b]
  \centering
  \caption{Hardware and software comparison of the smartphones we used in our cross-platform hologram stability experiment.}
  \vspace{-0.14in}
  \begin{tabular}{|P{3.35cm}|P{0.7cm}|P{1.0cm}|P{1.65cm}|}
  \hline
  \textbf{Device Model}&\textbf{RAM (GB) }&\textbf{Depth sensor}&\textbf{AR Platform}\\
  \hline
    iPhone 11&4&No&ARKit 4\\
    \hline
    Nokia 7.1&4&No&ARCore 1.23\\
    \hline
    Samsung Galaxy Note 10+&12&Yes&ARCore 1.23\\
    \hline
  \end{tabular}
  \label{tab:DeviceComparison}
\end{table}

Our results show that there is some consistency in terms of how challenging specific actions are for hologram stability in smartphone~AR. For example, the \textit{Unfocus} action, involving a relatively short device trajectory, no change in viewing position and no interruption to input data, consistently resulted in the lowest mean drift: 1.0cm for the iPhone, 2.5cm for the Nokia and 4.0cm for the Samsung. In contrast the \textit{Unfocused move} action, with the addition of a change in viewing position and a longer trajectory, resulted in \emph{over 10cm of drift for all devices}, namely 15.4cm for the iPhone, 12.6cm for the Nokia, and 10.1cm for the Samsung. However, we also find that there are important differences in hologram stability performance across platforms and devices, which we detail below.

\begin{figure}[]
\captionsetup[subfigure]{justification=centering}
\begin{subfigure}{.295\textwidth}
  \includegraphics[width=0.95\linewidth]{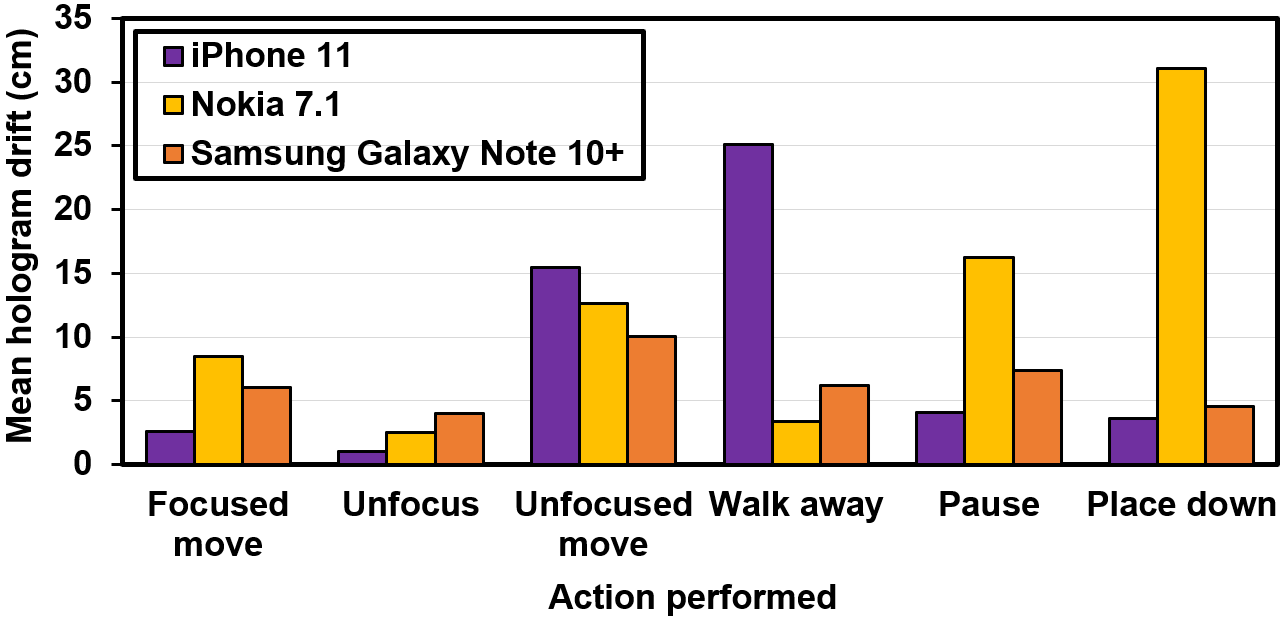}
  \vspace{-0.04in}
  \caption{Mean hologram drift}
  \label{fig:CrossPlatformDrift}
\end{subfigure}
\begin{subfigure}{.170\textwidth}
  \includegraphics[width=0.95\linewidth]{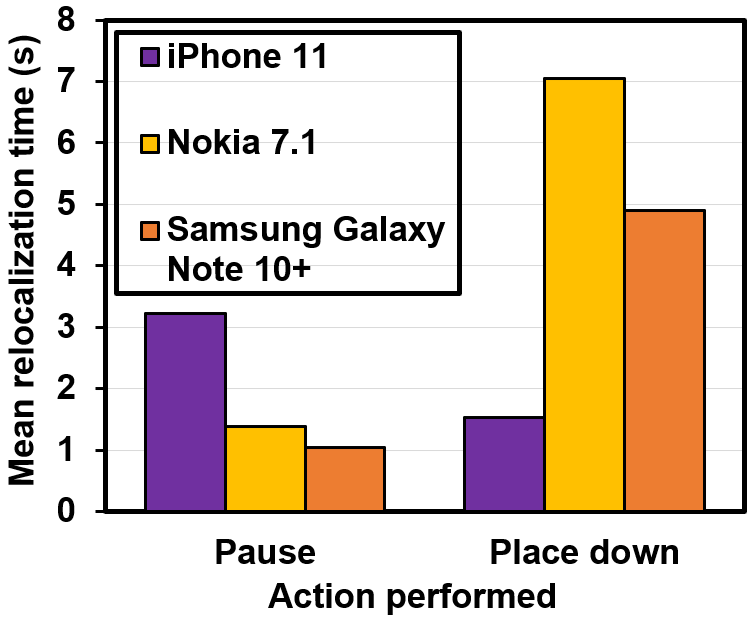}
  \vspace{-0.04in}
  \caption{Mean relocalization time}
  \label{fig:CrossPlatformReloc}
\end{subfigure}
\vspace{-0.1in}
\caption{Mean hologram drift and relocalization times for 3 devices after performing different actions in our cross-platform experiments. We observe significant differences between platforms and devices: for example, for \textit{Walk away}, mean drift ranges from 3.4cm on the Nokia 7.1 to 25.1cm on the iPhone~11.}
\label{fig:CrossPlatform}
\end{figure}

\textbf{ARKit vs. ARCore: }ARKit achieved lower mean hologram drift than ARCore for the majority of actions, and less than 0.5cm drift more frequently for all actions. As shown in Figure~\ref{fig:CrossPlatformDrift}, the mean drift was lower on ARKit than both ARCore devices for 4 out of 6 actions. Two of those actions were the less challenging \textit{Focused move} and \textit{Unfocus} actions; for \textit{Focused move} the mean drift on ARKit was 2.6cm and on ARCore 7.2cm (8.4cm on the Nokia and 6.0cm on the Samsung), and for \textit{Unfocus} the mean drift was 1.0cm on ARKit and 3.3cm on ARCore (2.5cm on the Nokia and 4.0cm on the Samsung). The other two actions were the actions that involved a temporary loss of tracking, \textit{Pause} and \textit{Place down}; for \textit{Pause} the mean drift on ARKit was 4.1cm and on ARCore 11.8cm (16.2cm on the Nokia and 7.4cm on the Samsung), and for \textit{Place down} the mean drift was 3.6cm on ARKit and 17.8cm on ARCore (31.1cm on the Nokia and 4.5cm on the Samsung).  

For one of the remaining two actions, \textit{Unfocused move}, drift was comparable, and if we remove one outlier scene in which drift was greater than 1m on ARKit, mean drift was actually lower on ARKit. The most notable exception was the \textit{Walk away} action, in which mean drift was much higher on ARKit, 25.1cm compared to an average of 4.8cm on ARCore (in fact ARCore outperformed ARKit in over 80\% of scenes). Interestingly, the ARCore devices temporarily lost tracking more frequently during the \textit{Walk away} action; the iPhone lost tracking in 35\% of scenes, for less than 1s each time, while the Nokia and the Samsung lost tracking in 80\% and 85\% of scenes respectively, often for multiple seconds at a time (0.2--11.0s on the Nokia, 0.1--9.2s on the Samsung). Yet once the user returned to view the space where the hologram was placed, ARCore was clearly able to achieve accurate loop closure (recognizing a previously visited place and adjusting the pose estimate accordingly) more reliably than ARKit, and drift that had accumulated throughout the action was usually corrected noticeably better. 

In terms of the relocalization time incurred by an interruption to sensor data, the results for ARKit and ARCore depended greatly on what type of interruption took place. The mean relocalization times for \textit{Pause} and \textit{Place down} on each device are shown in Figure~\ref{fig:CrossPlatformReloc}. On average the resumption of tracking after \textit{Pause} took longer on ARKit (3.2s) than on ARCore (1.2s). This indicates that ARCore can recover tracking more quickly after an interruption to both visual and inertial input data. However, for an interruption to visual data alone as in \textit{Place down}, ARKit regained tracking faster (1.5s) than ARCore on average (6.0s). The two results combined suggests that the VI-SLAM algorithms employed by each platform take different approaches to handling the resumption of tracking.

\textbf{Nokia 7.1 vs. Samsung Galaxy Note 10+: }Comparing the two Android smartphones, the Samsung, which was both released more recently and targeted at a higher price point, outperformed the Nokia in the majority of cases. For instance, significantly higher mean drift was observed on the Nokia~7.1 after the \textit{Pause} and \textit{Place down} actions than on the Samsung Galaxy Note 10+, 16.2cm compared to 7.4cm for \textit{Pause} and 31.1cm compared to 4.5cm for \textit{Place down}. The mean relocalization times for the \textit{Pause} and \textit{Place down} actions were also lower on the Samsung than the Nokia; for \textit{Pause} mean relocalization time was 1.4s on the Nokia and 1.0s on the Samsung, and for \textit{Place down} the mean relocalization time was 7.1s on the Nokia and 4.9s on the Samsung. This is likely due to a combination of the Samsung having greater computational resources, as well as the camera specifications of the two devices. The resolution of the camera images is not a factor (both devices have 12MP cameras, and the SLAM backend of ARCore appears to only use 640x480 pixel images), but the primary camera of the Samsung has two aperture modes ($f$/1.5 and $f$/2.4), which allows it to better adjust to different light levels, and the Nokia camera has a fixed aperture ($f$/1.8). Unlike the Nokia, the Samsung camera includes optical image stabilization~\cite{OpticalImageStabilization}, which should result in less blur and hence better feature detection, especially at low light levels. These differences in camera performance in darker conditions are reflected in our drift results; for environments in which the AR platform brightness estimation value was less than 0.4 (20\% of scenes), the mean drift across all actions was 19.6cm for the Nokia, but only 5.6cm for the Samsung. 


\emph{However, the Samsung Galaxy did not outperform the Nokia~7.1 for all actions}. For the \textit{Unfocus} action, mean drift was 2.5cm on the Nokia and 4.0cm on the Samsung, and drift was lower on the Nokia in 70\% of scenes. Higher drift was often recorded on the Samsung for black or reflective surfaces such as a stovetop, surfaces which are known to pose challenges for the type of time-of-flight depth sensor the phone is equipped with \cite{tofprinciples}. It was also higher in scenes in which the user was sitting down and the \textit{Unfocus} or the \textit{Unfocused move} action took the phone close to the floor; we observed in general that the Samsung camera often went out of focus temporarily when it came in close proximity with a surface. On the other hand, the addition of a depth sensor does appear to speed up plane detection significantly; the mean time to detection of the first plane was 2.8s compared to 7.9s on the Nokia (and 8.8s on the iPhone). This is certainly beneficial in terms of allowing the user to place holograms as quickly as possible, though we note that it has the potential to decrease hologram stability, because it generally means the time spent pre-mapping the environment is lower (for all of our 6 actions the mean mapping time was lower on the Samsung than the iPhone or the Nokia, by approximately 25\%); in a shorter mapping period the user generally moves around less, and maps less of the surrounding environment.

\subsection{ARKit Visual Environments Experiments}
\label{subsec:ARVisualEnvironmentsDataset}
\begin{figure}[]
\includegraphics[width=1\linewidth]{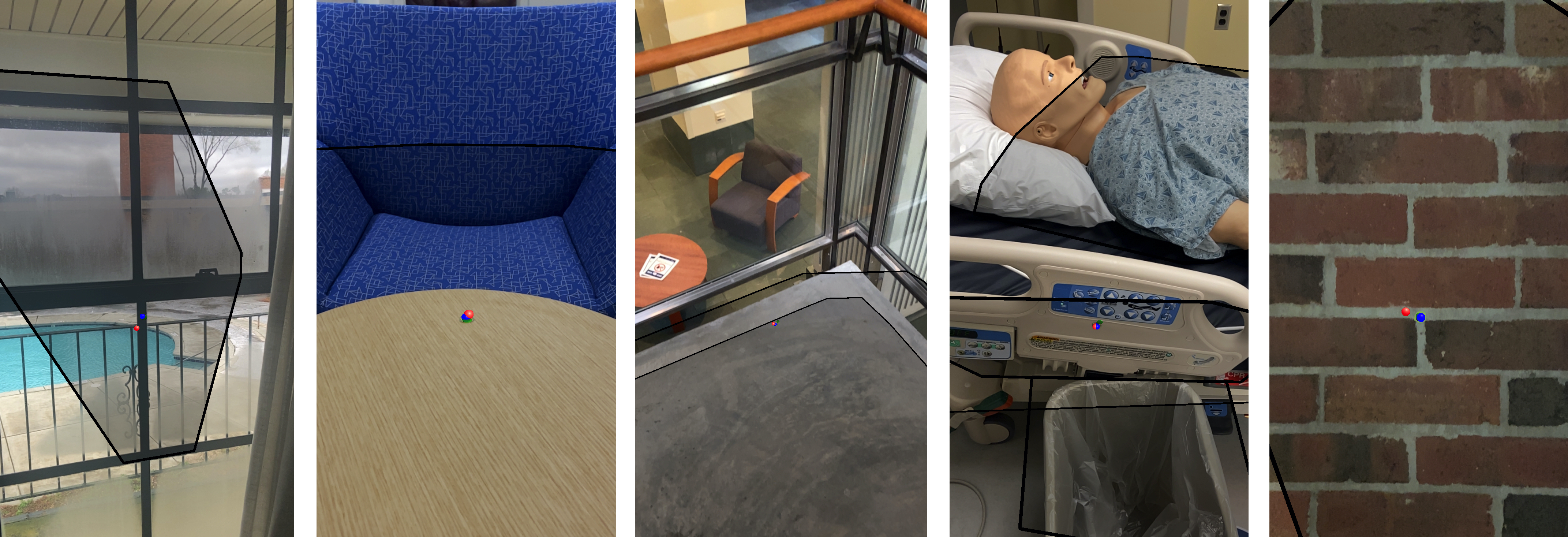}
\vspace{-0.24in}
\caption{Examples of the settings we performed measurements in for the ARKit visual environments experiments (Section~\ref{subsec:ARVisualEnvironmentsDataset}).}
\label{fig:VisualEnvironments}
\end{figure}

Next we address our goal of measuring hologram stability in a much wider variety of visual conditions. Because ARKit achieved lower drift for the majority of the actions we tested in Section~\ref{subsec:CrossPlatformHologramStabilityExperiment}, we chose this platform for our evaluation, and conducted all trials on an iPhone~11 running iOS 14.4 and ARKit 4. We captured over 200 scenes for each of the 6 actions, 1363 scenes in total.\footnote{The dataset we collected is available upon request.} These scenes span 21 different rooms in 5 different buildings, including a living room, bedroom, kitchen and bathroom in a small apartment, labs, offices and corridors in large academic buildings at Duke University, as well as the Health Innovation Lab \cite{HealthInnovationLab} at Duke University School of Nursing, a simulated clinical environment for teaching and product development. We sought out and created a diverse set of visual environments that might host AR, including repetitive fabrics and brickwork, a glass-walled meeting room, a multi-storey lobby, and an ICU simulation room. Examples of some of the environments are shown in Figure~\ref{fig:VisualEnvironments}. \emph{To the best of our knowledge, this study is the first to accurately measure AR virtual content positional errors in a large number of different environments, incorporating a wide range of visual conditions.}

\begin{figure}
\captionsetup[subfigure]{justification=centering}
\begin{subfigure}{.23\textwidth}
  \includegraphics[width=1\linewidth]{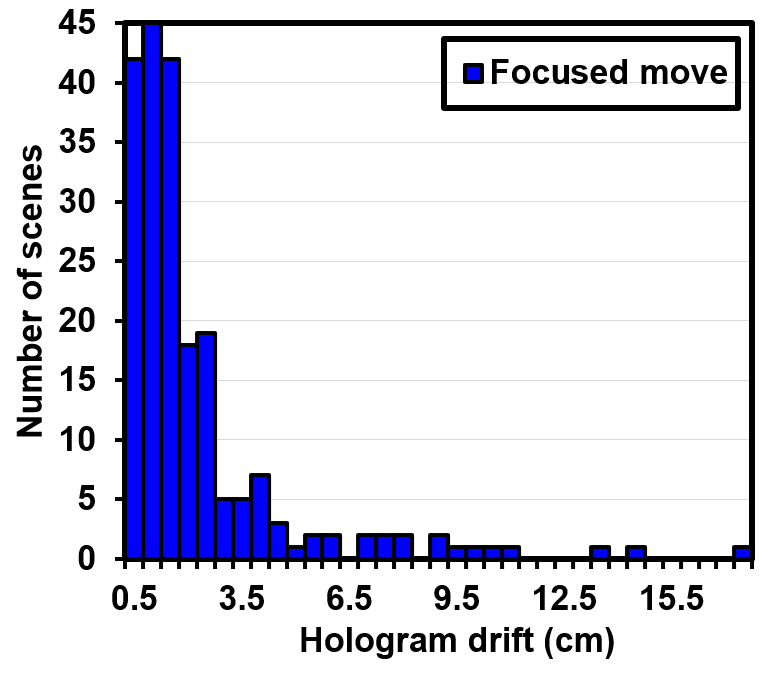}
  \vspace{-0.2in}
  \caption{}
  \label{fig:ARKitDriftHistogramFM}
\end{subfigure}
\begin{subfigure}{.23\textwidth}
  \includegraphics[width=1\linewidth]{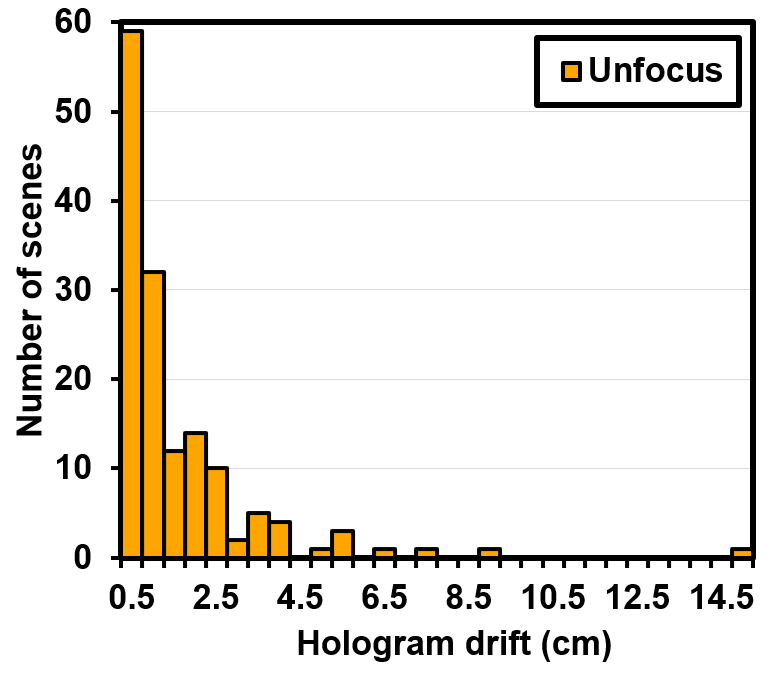}
  \vspace{-0.2in}
  \caption{}
  \label{fig:ARKitDriftHistogramU}
\end{subfigure}
\begin{subfigure}{.23\textwidth}
  \includegraphics[width=1\linewidth]{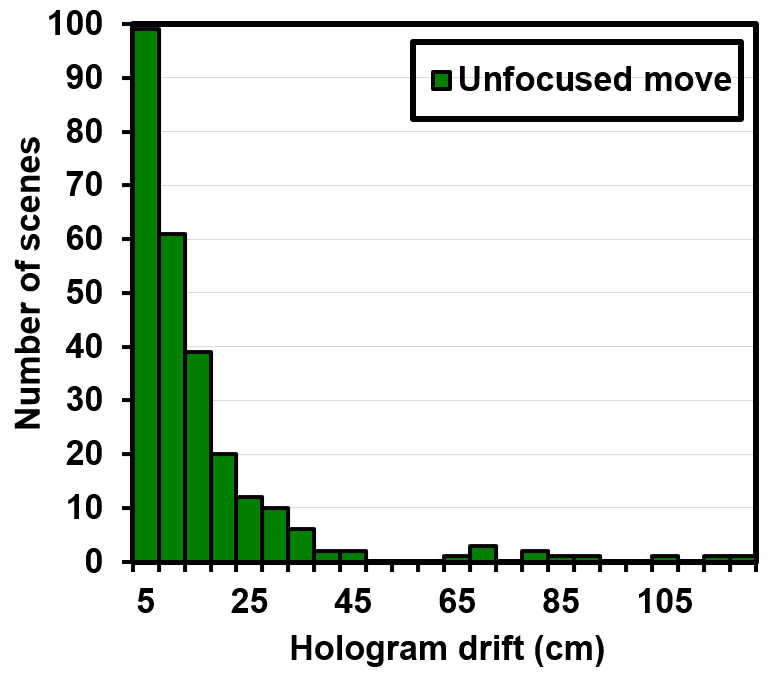}
  \vspace{-0.2in}
  \caption{}
  \label{fig:ARKitDriftHistogramUM}
\end{subfigure}
\begin{subfigure}{.23\textwidth}
  \includegraphics[width=1\linewidth]{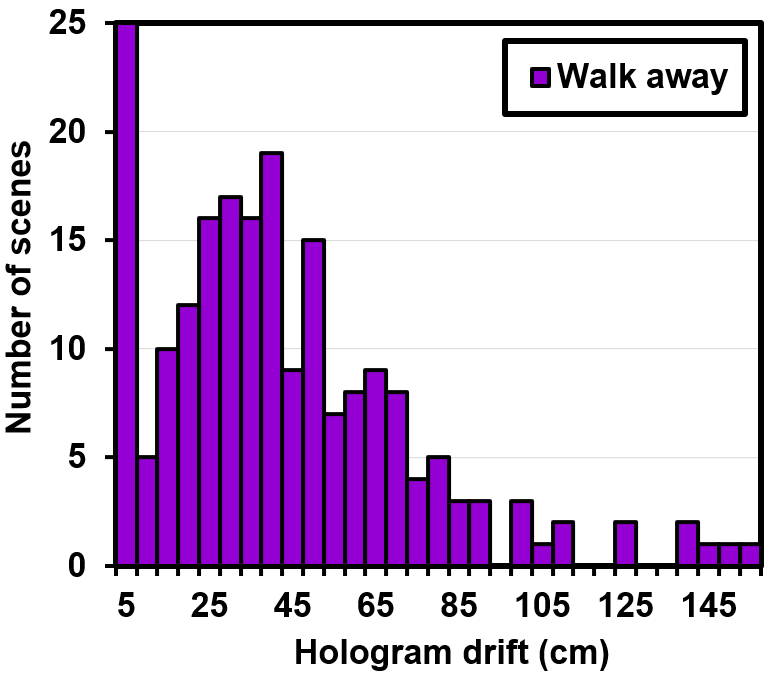}
  \vspace{-0.2in}
  \caption{}
  \label{fig:ARKitDriftHistogramWA}
\end{subfigure}
\begin{subfigure}{.23\textwidth}
  \includegraphics[width=1\linewidth]{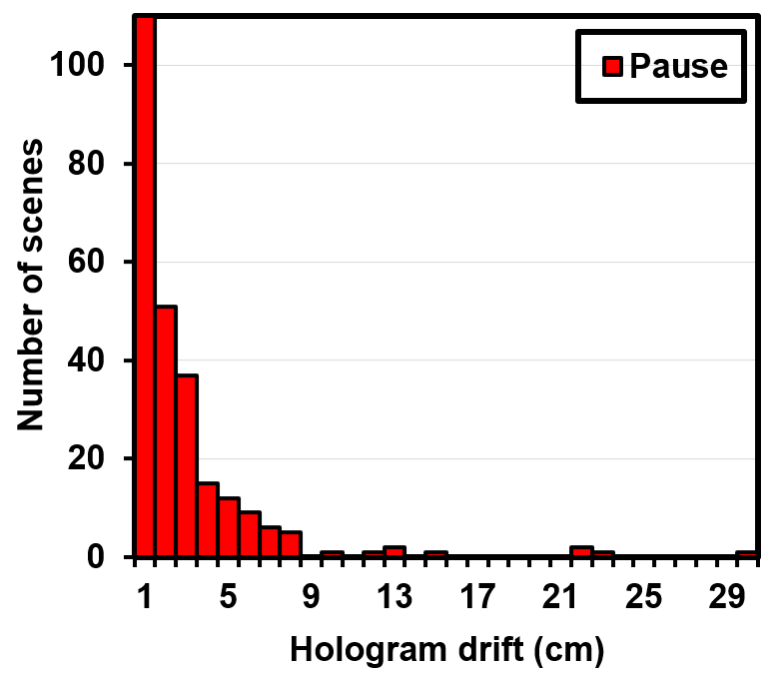}
  \vspace{-0.2in}
  \caption{}
  \label{fig:ARKitDriftHistogramP}
\end{subfigure}
\begin{subfigure}{.23\textwidth}
  \includegraphics[width=1\linewidth]{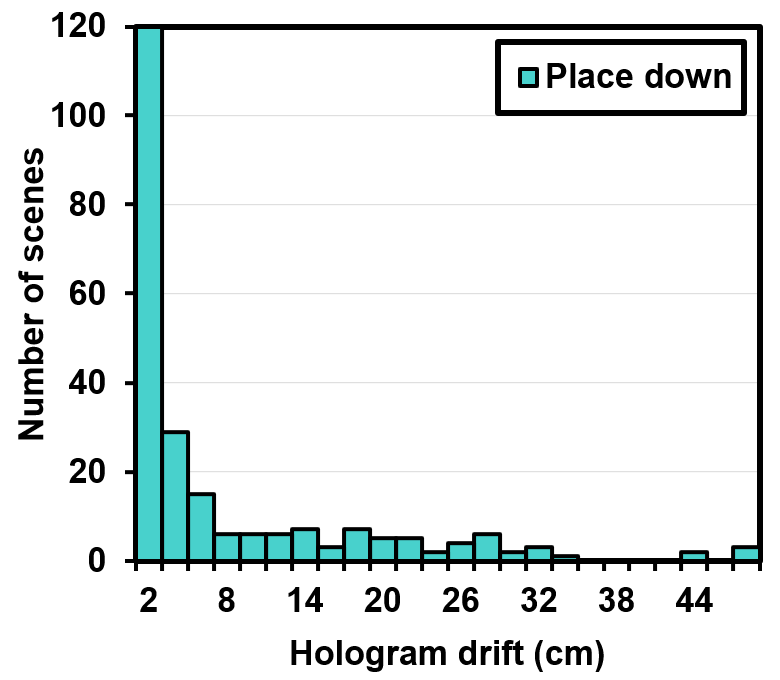}
  \vspace{-0.2in}
  \caption{} 
  \label{fig:ARKitDriftHistogramPD}
\end{subfigure}
\vspace{-0.12in}
\caption{Histograms of hologram drift on ARKit after performing 6 different actions (described in Section~\ref{subsec:DeviceMovementinSmartphoneAR}): \textit{Focused move} (a), \textit{Unfocus} (b), \textit{Unfocused move} (c), \textit{Walk away} (d), \textit{Pause} (e), and \textit{Place down} (f). All drift histograms have peaks at small drift values, though larger values are much more common for the \textit{Walk away} action.}
\label{fig:ARKitDriftHistograms}
\end{figure}

\begin{figure}
\captionsetup[subfigure]{justification=centering}
\begin{subfigure}{.23\textwidth}
  \includegraphics[width=1\linewidth]{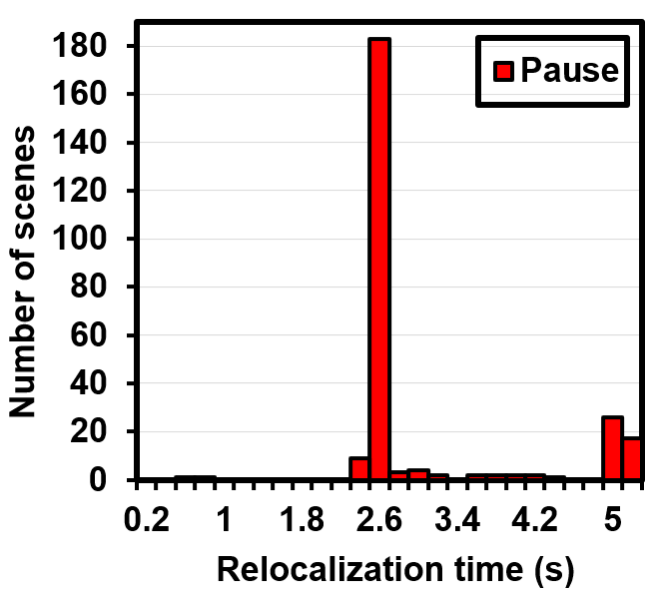}
  \vspace{-0.2in}
  \caption{}
  \label{fig:ARKitRelocP}
\end{subfigure}
\begin{subfigure}{.23\textwidth}
  \includegraphics[width=1\linewidth]{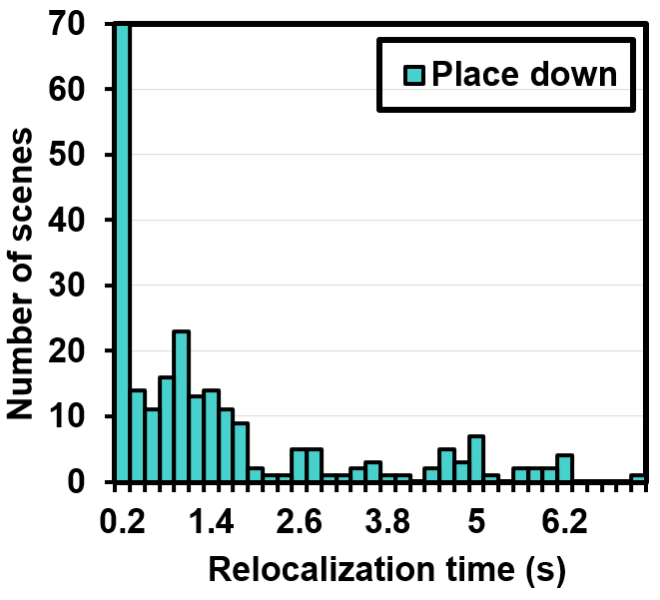}
  \vspace{-0.2in}
  \caption{}
  \label{fig:ARKitRelocPD}
\end{subfigure}
\vspace{-0.12in}
\caption{Histograms of relocalization time on ARKit after performing 2 actions, \textit{Pause} (a) and \textit{Place down} (b) (described in Section~\ref{subsec:DeviceMovementinSmartphoneAR}) which involve an interruption to pose tracking input data.}
\label{fig:ARKitRelocHistograms}
\end{figure}

The statistics on hologram drift for our 6 actions are shown in Table~\ref{tab:DriftStatistics}, and drift histograms in Figure~\ref{fig:ARKitDriftHistograms}.\footnote{The presented figures exclude the following outliers: 3 drift values greater than 160cm for \textit{Walk away} in Figure~\ref{fig:ARKitDriftHistogramWA}: 186.5cm, 188.2cm and 323.3cm; 1 drift value greater than 30cm for \textit{Pause} in Figure~\ref{fig:ARKitDriftHistogramP}: 44.9cm; 1 drift value greater than 50cm for \textit{Place down} in Figure~\ref{fig:ARKitDriftHistogramPD}: 82.2cm.} The statistics on relocalization time for the two actions which caused interruptions to sensor data (\textit{Pause} and \textit{Place down}) are shown in Table~\ref{tab:RelocStatistics}, and relocalization time histograms in Figure~\ref{fig:ARKitRelocHistograms}. Below we examine our results for each action in turn. For all actions we observed a strong negative effect on hologram stability at low light levels, when the brightness estimation value provided by the AR platform was less than 0.3. Clearly a well-lit environment is a key priority for those wishing to render stable holograms. However there are a number of other important factors to consider, as we now discuss.

\textbf{Focused move: }The \textit{Focused move} action involves a slow, controlled movement, with the user keeping the camera focused on the region of space where the hologram was placed. It resulted in one of the lowest values for mean hologram drift, 2.0cm, with drift of less than 0.5cm in 20\% of scenes, while large drift of greater than 10cm only occurred in 2\% of scenes. The mean estimated trajectory distance (calculated from the device position estimates for each frame) was 0.6m, but unlike some of the other actions we performed, this was dependent on how far away from the camera the hologram was originally placed; in order to change their viewing angle by $45^{\circ}$, a user has to move farther the farther away a hologram is. For this action the Spearman correlation coefficient between drift and trajectory distance was 0.5, and drift and hologram placement distance 0.4, illustrating how drift tends to be greater the farther away a hologram is placed, and the subsequent distance traveled before viewing the hologram again. Another reason for greater instability at greater placement distances is that the detection of recognizable visual features used for tracking is less accurate and consistent at greater viewing distances.

In general, large drift errors in the \textit{Focused move} scenes were the result of poor plane alignment. If the virtual plane a hologram is placed on is not well-aligned with the real world plane which the real reference point is placed on (due to an inaccurate point cloud generated while mapping), a different viewing angle may reveal this misalignment. A hologram then appears to drift out of position relative to the real reference point, even if it does not move in world space. We found that the two most common culprits for poor plane alignment were blank, featureless walls and doors, along with uneven surfaces such as objects on a table, where the height of the detected plane corresponded to the height of those objects, rather than the portion of the table upon which the hologram was placed. ARKit 4 introduced plane detection aided by machine learning, which speeds up plane detection and enables detection over textureless regions, aiding hologram placement. However, our experiments show that this comes at a price in terms of hologram stability; high drift often occured in these regions, characterized by camera images with low edge strength (measured using variance of the Laplacian as in \cite{garforth2019visual}), and point clouds with low density (less than 50 points per cubic meter).

\textbf{Unfocus: }Consistent with the results in our cross-platform experiments (Section~\ref{subsec:CrossPlatformHologramStabilityExperiment}), mean drift was lowest for the \textit{Unfocus} action, at~1.2cm. Drift was less than 0.5cm in 41\% of scenes, and greater than 10cm in only 1\% of scenes. While the movement associated with this action is more challenging (higher accelerometer and gyroscope values) than in \textit{Focused move}, and the mean estimated trajectory was longer (1.9m), in this action the user views the hologram from the same position. However, we again observed the trend that high drift was associated with low edge strength in camera images or low density in the generated point cloud. There are also cases in which a challenging environment to the right of the user conducting these experiments appears to have caused tracking errors; this is because the user was right-handed, and when a right-handed user lets the phone fall to their side in this action the camera faces to the right. The same would apply to the environment to the left for a left-handed user. In our experiments we note several high-drift scenes in which reflective elevator doors, long empty corridors, featureless walls or a repetitive texture such as a metal grid were to the right of the user. \emph{This highlights the importance of considering the entire visual environment when assessing its suitability for AR, as a user's device may face in unexpected directions.} 
    
\begin{table}[]
  \centering
  \caption{Hologram drift statistics for each of the actions we performed after hologram placement, in our ARKit visual environments experiments (SD = Standard Deviation).}
  \vspace{-0.14in}
  \begin{tabular}{|P{2.05cm}|P{1.2cm}|P{0.7cm}|P{0.7cm}|P{0.7cm}|P{0.7cm}|}
  \hline
\multirow{2}{2.05cm}{\centering \textbf{Action performed}} & \multirow{2}{1.2cm}{\centering \textbf{Number of scenes}} & \multicolumn{4}{c|}{\textbf{Drift (cm)}}\\
\cline{3-6}
 && \textbf{Mean} & \textbf{Min} & \textbf{Max} & \textbf{SD}\\
\hline
    \textit{Focused move}&205&2.0&0.1&17.1&2.5\\
    \hline
    \textit{Unfocus}&201&1.2&0.1&14.9&1.6\\
    \hline
    \textit{Unfocused move}&262&12.8&0.1&118.3&17.5\\
    \hline
    \textit{Walk away}&207&43.2&0.1&323.3&39.1\\
    \hline
    \textit{Pause}&255&2.5&0.0&44.9&4.4\\
    \hline
    \textit{Place down}&233&7.1&0.0&82.2&11.1\\
    \hline
  \end{tabular}
  \label{tab:DriftStatistics}
\end{table}

\textbf{Unfocused move: }Incorporating both the more challenging device movement in \textit{Unfocus} and the change in viewing angle in \textit{Focused move}, \textit{Unfocused move} resulted in much higher drift than either, in fact the second-highest mean drift of our 6 actions, at~12.8cm. Just 1\% of scenes resulted in a drift of less than 0.5cm, while \emph{greater than 10cm of drift was observed in 39\% of scenes}. The mean estimated trajectory distance was 2.8m, and there was an even stronger Spearman correlation coefficient between drift and estimated trajectory distance, 0.7, than in \textit{Focused move}. This is in part due to longer trajectories resulting in larger errors, but a long estimated trajectory was also sometimes the result of large adjustments in estimated device position between frames. For the \textit{Unfocused move} scenes spikes in the inter-frame estimated trajectory distance were a good indicator of hologram instability; no scenes in which drift was less than 1.5cm incurred a single inter-frame distance of greater than 5cm.

We find that increased complexity in the camera images sampled during mapping is beneficial for tracking, particularly for this action. We measure complexity using Shannon entropy \cite{shannon1948mathematical}, and the entropy $E$ of an image $I$ for an 8-bit grayscale image is then $E(I) \in [0,8]$. We take the minimum value from the sampled camera images, and for the \textit{Unfocused move} scenes the range of entropy values was 2.2--7.8, but for all scenes in which drift was less than 2cm, entropy was greater than 6.9. However, as with the \textit{Unfocus} action, we saw evidence that featureless walls or doors and long corridors to the right of the user, visible to the camera when the smartphone is dropped to their side, can still cause tracking loss and high drift. In one scene a hologram was placed on a pillowcase with a distinctive pattern (the minimum entropy in the camera images sampled during mapping was 7.25), but there was a white door approximately 1.5m to the right of the user. During the time the smartphone was facing in that direction tracking was lost for 0.3s, and 63cm of drift was recorded for the scene.

\textbf{Walk away: }By far the longest mean estimated trajectory for our actions was for \textit{Walk away}, at 14.7m, and this action also resulted in by far the highest mean drift, 43.2cm. 5\% of scenes resulted in less than 0.5cm of drift, while the vast majority, 86\%, resulted in drift of greater than 10cm, and 67\% resulted in drift of greater than 25cm. Because of the fact that pose tracking error tends to increase with distance traveled, we observe that by the time the user returns to their original viewpoint after this trajectory the pose estimate is invariably incorrect. Therefore for the hologram to appear in the correct position the VI-SLAM algorithm must recognize a place that had previously been viewed from the camera images, and adjust the pose estimate accordingly. In our examination of the inter-frame estimated trajectory distance for each scene we see when this occurs in the form of a spike related to the position adjustment, at the end of the user's movement. All \textit{Walk away} scenes in which drift was less than 3cm have these position adjustment spikes where inter-frame distance is greater than 5cm, showing how crucial place recognition and loop closure are to hologram stability for this action. Scenes in which place recognition occurred are characterized by a minimum AR platform brightness estimate of 0.4, and a minimum of 500 FAST corners \cite{rosten2006machine} detected in each sampled camera image.

\textbf{Pause: }The \textit{Pause} action resulted in an interruption to both visual and inertial input data, which is challenging for any pose tracking system. However, the mean drift for this action was only 2.5cm, with 28\% of scenes resulting in less than 0.5cm drift and just 4\% of scenes resulting in greater than 10cm of drift. This relatively low level of drift is due to there being little device movement during the interruption (the mean estimated trajectory length was 0.5m, though this is an overestimate due to a few large values associated with device position estimate errors). Due to the interruption to input data, this action also incurred a period of relocalization; the mean relocalization time was 2.9s. The histogram of relocalization times for the \textit{Pause} action is shown in Figure~\ref{fig:ARKitRelocP}. Unlike all other drift and relocalization time histograms in our experiment the distribution is bimodal, with relocalization completed in approximately 2.5s in~75\% of scenes, but taking approximately 5s in 17\% of scenes.


Both drift and relocalization time increased at low brightness levels as expected. However, the Spearman correlation coefficient between drift and relocalization time was only 0.3, indicating that aside from lighting conditions, it is not the same factors which cause these different types of hologram instability. Drift tended to be greater when holograms were placed farther away from the device; the mean placement distance of scenes with less than 0.5cm drift was 0.65m, but in scenes with greater than 5cm drift it was 1.05m. High drift also sometimes occurred when there was low contrast or low edge strength in the camera images; this is a reflection of the fact that accurate relocalization requires distinguishable visual features.

\begin{table}[]
  \centering
  \caption{Relocalization time statistics for the actions we performed which caused a loss of tracking through interruptions to sensor data, in our ARKit visual environments experiments (SD = Standard Deviation).}
  \vspace{-0.14in}
  \begin{tabular}{|P{2.05cm}|P{1.2cm}|P{0.7cm}|P{0.7cm}|P{0.7cm}|P{0.7cm}|}
  \hline
\multirow{2}{2.05cm}{\centering \textbf{Action performed}} & \multirow{2}{1.2cm}{\centering \textbf{Number of scenes}} & \multicolumn{4}{c|}{\textbf{Relocalization time (s)}}\\
\cline{3-6}
 && \textbf{Mean} & \textbf{Min} & \textbf{Max} & \textbf{SD}\\
\hline
    \textit{Pause}&255&2.9&0.5&5.1&1.0\\
    \hline
    \textit{Place down}&233&1.4&0.0&7.1&1.7\\
    \hline
  \end{tabular}
  \label{tab:RelocStatistics}
\end{table}

\textbf{Place down: }The \textit{Place down} action involves significant device movement, when the user places their smartphone down on a nearby surface (the mean estimated trajectory length was 2.7m), as well as an interruption to the visual input data, because the rear camera is facing downward onto that surface. The mean drift for this action was 7.1cm, with less than 0.5cm of drift in 23\% of scenes, and greater than 10cm of drift in 25\% of scenes. The mean drift was higher than for the \textit{Pause} action because of the additional device movement, but the mean relocalization time, 1.4s, was lower than \textit{Pause} because the \textit{Place down} action does not interrupt the inertial data input. Relocalization was completed in 2s or less in 79\% of scenes.

We observe that there is a consistent pattern in the inter-frame estimated trajectory distance for this action. One spike always occurs after the phone is placed down, which corresponds to a pose tracking error when the camera is covered. Then in the vast majority of cases a large position adjustment also occurs when the motion of picking the phone up has been completed and the camera view is recognized. We find evidence that low contrast and edge strength are correlated with higher drift scenes; for example the mean contrast and variance of the laplacian for all \textit{Place down} scenes were 58 and 51, but the mean values for scenes with 10cm of drift or greater were lower, 48 and 34 respectively. This again indicates how a well-defined texture in the region of space where a hologram is placed is advantageous for accurate place recognition and relocalization, necessary after an interruption to the camera image inputs (as in \textit{Pause} or \textit{Place down}), or a larger movement away from the area (as in \textit{Walk away}).

\vspace{0.1in}

\begin{table}[b]
  \centering
  \caption{Hologram drift statistics for each of the actions we performed after hologram placement, in well-lit and textured environments, in our ARKit visual environments experiments (SD = Standard Deviation).}
  \vspace{-0.14in}
  \begin{tabular}{|P{2.05cm}|P{1.2cm}|P{0.7cm}|P{0.7cm}|P{0.7cm}|P{0.7cm}|}
  \hline
\multirow{2}{2.05cm}{\centering \textbf{Action performed}} & \multirow{2}{1.2cm}{\centering \textbf{Number of scenes}} & \multicolumn{4}{c|}{\textbf{Drift (cm)}}\\
\cline{3-6}
 && \textbf{Mean} & \textbf{Min} & \textbf{Max} & \textbf{SD}\\
\hline
    \textit{Focused move}&139&1.4&0.1&10.0&1.5\\
    \hline
    \textit{Unfocus}&123&0.9&0.1&5.3&1.0\\
    \hline
    \textit{Unfocused move}&186&12.4&0.1&118.3&17.8\\
    \hline
    \textit{Walk away}&137&39.9&0.1&323.3&37.5\\
    \hline
    \textit{Pause}&138&1.8&0.0&13.0&2.1\\
    \hline
    \textit{Place down}&169&5.2&0.0&46.2&8.5\\
    \hline
  \end{tabular}
  \label{tab:RecommendedDriftStatistics}
\end{table}

\begin{table}[b]
  \centering
  \caption{Relocalization time statistics for the actions we performed which caused a loss of tracking through interruptions to sensor data, in well-lit and textured environments, in our ARKit visual environments experiments (SD = Standard Deviation).}
  \vspace{-0.14in}
  \begin{tabular}{|P{2.05cm}|P{1.2cm}|P{0.7cm}|P{0.7cm}|P{0.7cm}|P{0.7cm}|}
  \hline
\multirow{2}{2.05cm}{\centering \textbf{Action performed}} & \multirow{2}{1.2cm}{\centering \textbf{Number of scenes}} & \multicolumn{4}{c|}{\textbf{Relocalization time (s)}}\\
\cline{3-6}
 && \textbf{Mean} & \textbf{Min} & \textbf{Max} & \textbf{SD}\\
\hline
    \textit{Pause}&138&2.9&2.3&5.1&1.0\\
    \hline
    \textit{Place down}&169&1.2&0.0&6.1&1.5\\
    \hline
  \end{tabular}
  \label{tab:RecommendedRelocStatistics}
\end{table}

\subsection{Recommended ARKit Visual Environments}
\label{subsec:RecommendedARKitVisualEnvironments}
The results presented in the previous subsection cover all of the visual environments in which we performed hologram stability measurements on ARKit. We now consider only the visual environments which were well-lit and sufficiently textured, in accordance with the ARKit and ARCore guidelines \cite{ARKitGuidelines,ARCoreGuidelines}, which provide descriptions of conditions that may hinder accurate detection of the real world environment. After testing a variety of different metrics, we used the following parameters to define these conditions: a light level (brightness) estimate from the AR platform of 0.3 or greater, a minimum variance of the Laplacian (edge strength) of 20 in each of the sampled camera images (this metric was used to measure camera image texture for visual-only SLAM in \cite{garforth2019visual}), and a point cloud density (generated during mapping prior to hologram placement) of at least 50 points per cubic meter.  

The drift and relocalization time statistics for each each action are shown in Table~\ref{tab:RecommendedDriftStatistics} and Table~\ref{tab:RecommendedRelocStatistics} respectively. We can see that compared to mean drift in all environments (Table~\ref{tab:DriftStatistics}), mean drift was lower by at least 25\% for the \textit{Focused move}, \textit{Unfocus}, \textit{Pause} and \textit{Place down} actions. However for the \textit{Unfocused move} and \textit{Walk Away} actions, which have the longest trajectories, and the most camera frames covering regions of the environment not mapped before hologram placement, there was only a minor reduction in mean drift (3\% for \textit{Unfocused move} and 8\% for \textit{Walk Away}). There was also little impact on relocalization time, with the expected value for \textit{Pause} remaining the same and the expected value for \textit{Place down} lower by only 0.2s (14\%). Overall, \emph{noticeable drift was still present if we select environments that should minimize it}, and in many scenarios, hologram stability is not improved significantly. We experimented with multiple definitions of `good scenes', subjective and objective, and found this to consistently be the case.

\section{Discussion}
\label{sec:Discussion}

In our study of markerless smartphone AR, \emph{we observed measurable holographic drift for all devices and all user actions we examined}, with the expected drift ranging from 1.0--4.0cm for the `easy' \emph{Unfocus} action to as high as 25.1cm and 31.1cm for actions that are more `challenging' for VI-SLAM (\emph{Walk away} on iPhone 11, \emph{Place down} on Nokia 7.1). It is clear that in these scenarios placed holograms are not yet `here to stay', and that high-precision applications (e.g., medical or industrial scenarios) are not yet supported in markerless~AR on smartphones. 

\par \noindent 
\textbf{Mobile platform comparison}: Our direct comparison of hologram stability on 3 different mobile phone models with 2 different AR platforms in identical environments (Section~\ref{subsec:CrossPlatformHologramStabilityExperiment}) revealed considerable differences in both drift and relocalization time across different devices, as could be expected due to the differences in cameras, inertial sensors, and computational capabilities of different smartphones. We noted significant differences between ARKit and ARCore: for instance, we observed ARKit associated with lower drift for most actions but not for \emph{Walk away}, and observed it resuming tracking faster than ARCore for the \emph{Place down} action, but not for \emph{Pause}. This implies that \emph{while cross-platform tools such as Unity's AR Foundation~\cite{ARFoundation} facilitate the development of apps for multiple platforms, virtual content stability testing needs to be conducted separately on different platforms}. Curiously, our results also showed that a higher-end smartphone (the Samsung Galaxy Note 10+) did not always out-perform a mid-range one (the Nokia 7.1); this may indicate that developers cannot rely on testing a lower-end device to establish a minimum level of hologram stability performance (especially when devices may be equipped with different types of environment sensors). 



\par \noindent 
\textbf{Holographic drift in simple scenarios}: In our experiments with a wide variety of visual environments with ARKit (Sections~\ref{subsec:ARVisualEnvironmentsDataset} and~\ref{subsec:RecommendedARKitVisualEnvironments}), over a short period of time, when a user gets briefly distracted but then views the hologram from the same position (\emph{Unfocus}), the expected drift was only 1.2cm, and was less than 0.5cm in 40\% of the scenes. While noticeable if the hologram is placed in a well-recognizable position in the real world (such as the real reference point we used in our experiments), this is frequently not noticeable in natural environments, and is likely to not be distracting for the users in many applications. Restricting the usage of AR to well-lit and well-textured conditions, as is recommended by both the ARKit and ARCore developer guidelines \cite{ARKitGuidelines,ARCoreGuidelines}, further reduces the expected drift to below 1cm. Overall, this indicates that holograms designed for viewing or interaction from one side will generally appear stable (e.g., in 2D text or video, games such as Pokemon GO~\cite{PokemonGo} or Angry Birds AR \cite{AngryBirds}, or quick representations of product sizes as in IKEA Place \cite{IKEAPlace} or Amazon AR View \cite{AmazonARview}). 

The expected drift increases when the user changes, even relatively slightly, her position with respect to the hologram (\emph{Focused move} action, 45$^{\circ}$ viewing angle change). In this case, mean drift increases 
to 2.0cm, with 80\% of the scenes resulting in drift of more than 0.5cm. This indicates that 
when holograms are designed to appear like real, complex 3D objects for users to explore in detail from multiple angles (e.g., objects in a virtual museum, a high-quality 3D representation of a product), spatial stability of the holograms becomes much more of an issue. Restricting the usage of AR to well-lit and well-textured conditions 
does help reduce the expected drift to more manageable levels, namely from 2.0cm to 1.4cm in our experiments. 
We note, however, that these restrictions do not always feel natural. In particular, \emph{many surfaces one would commonly encounter in normal conditions are actually `bad' scenes due to there being no high-quality texture on them} (e.g., tables without a clearly distinguishable surface pattern or a object on them that provided additional texture were a common cause of high drift in our experiments). Even in such a simple scenario, there is room for improvement in hologram stability. 


\par \noindent 
\textbf{Holographic drift  with increased user mobility}: Drift increases significantly when the user performs other actions, for example letting the phone drop to their side then changing their viewing position (\emph{Unfocused move}; expected drift 12.8cm), or walking away from the scene and returning to view it from the same angle (\emph{Walk away}; expected drift 43.2cm).\footnote{We note that in our cross-platform experiments  (Section~\ref{subsec:CrossPlatformHologramStabilityExperiment}), for ARCore, \emph{Unfocused move} was associated with a high drift, but \emph{Walk away} was not. For ARCore, the level of drift associated with the \emph{Walk away} action was similar to the level of drift associated with the \emph{Focused move} action. Additional studies are required to assess the performance of ARCore on the \emph{Walk away} action across a wider range of environmental conditions.} For \emph{Unfocused move}, drift exceeded 0.5m in many cases (see Figure~\ref{fig:ARKitDriftHistogramUM}); for \emph{Walk away}, there were multiple instances of drift exceeding 1m (see Figure~\ref{fig:ARKitDriftHistogramWA}). This magnitude of error is large and obvious, and may make an AR experience unnatural, frustrating, and potentially unusable; an example of this is shown in Figure~\ref{fig:Drift2}, in which a user has walked away, possibly to get another person to show the virtual IKEA lamp to, but when they return the lamp is not on the table where they left it. High instability is associated with cases where the user is active; pose estimation error accumulates over longer trajectories, and if accurate place recognition and loop closure are not achieved when returning to a hologram, then high hologram drift is frequently observed. This problem is exacerbated on smartphones compared to headsets for several reasons, including two related to camera views. Firstly, a handheld device is often by a user's side when the user moves, in which case the camera view may easily be obstructed or challenging due to the close proximity of walls or low light areas (on a headset cameras are always facing in the direction the user is facing). Secondly, smartphone form factor only supports environment-facing cameras angled in one direction, which limits their field of view (FoV); while the FoV of headset cameras can likely extend beyond one textureless region, a phone's is more likely to be limited to it. Further enhancements to SLAM algorithms, specific to the challenges of handheld devices, are needed to address this problem, if smartphone AR is to support 
experiences that involve significant user movement and exploration of a space.



\par \noindent 
\textbf{Drift and relocalization artifacts with AR experience interruptions}: We also studied two actions associated with the interruption of the AR experience for a brief period of time, which is known to cause problems (ARKit guidelines~\cite{ARKitGuidelines} suggest that developers minimize the number of times users have to exit their AR experience to perform a task outside of AR). We specifically examined the case of one 5s-long interruption, in which the user pauses an app but does not change the position of the phone by much (\emph{Pause}) and one in which the user stops using the phone and places it down, temporarily covering the rear camera, without pausing the app first (\emph{Place down}). In these cases we observe significant drift (on average 2.5cm for \emph{Pause} and 7.1cm for \emph{Place down}); we also observe prolonged periods of relocalization (on average 2.9s for \emph{Pause}, 1.4s for \emph{Place down}, and up to 5.1s for \emph{Pause} and 7.1s for \emph{Place down}). During these relocalization periods holograms are in a high state of instability, with holograms frequently appearing in the wrong position, hologram movement observable even while the device is stationary, and some hologram `flickering' (rapid disappearance and reappearance of the hologram). 
The impact of 
interruptions on hologram stability is problematic for AR experiences designed to last longer than a few minutes, as it becomes more and more likely over time that a user will want to perform a task outside of AR, such as send a text message, write something down, or grab a drink. Again compared to headsets, the cameras of which are not usually covered if the device is placed down during use, 
developing VI-SLAM techniques which can achieve relocalization in an accurate and timely manner takes on even more importance in smartphone AR.

\par \noindent 
\textbf{Inherent challenges for markerless smartphone AR in typical indoor environments}: Our experiments, conducted in a wide range of visual environments, underscore a critical property of pose calculations in VI-SLAM-based markerless mobile AR systems: \emph{it is not just the properties of the surface the hologram is placed on that matter; surrounding surfaces have an impact as well, because a user's camera often points in other directions between hologram views}. It is a fundamental challenge for markerless AR that to achieve hologram stability, pose estimation error needs to be low (ideally, sub-cm), but textureless areas in particular are known to be problematic~\cite{jinyu2019survey}, and textureless areas are ubiquitous in built environments. \emph{While it is relatively straightforward to ensure that one specific surface is well-lit, detailed, and not reflective, it is significantly harder to ensure that an entire environment is like this}. Indeed, the frequency of challenging camera views is often high in smartphone AR because of the limited FoV of environment-facing cameras, and movement patterns of a handheld device. 
In ongoing work we are developing approaches that use data from other AR sessions and external cameras to capture the properties of the entire environment, towards more accurate predictions of hologram stability. Much work remains in fully characterizing how visual conditions and device movement affect performance, but this is an exciting direction which will inform AR app design, the design of spaces that host AR, and the development of new VI-SLAM techniques to tackle common problems.

\smallskip  
\par \noindent \textbf{Limitations}: The presented study examines hologram drift and relocalization time, but not the other types of virtual content positioning errors commonly observed in AR, such as instability of hologram orientation, hologram jitter (high frequency shaking), and hologram judder (uneven motion)~\cite{HologramStability}. 
    In addition, the study does not examine the interplay of objective hologram stability metrics and subjective user experience, which is likely to be non-trivial, with user perception of hologram stability influenced by multiple factors including the contrast between the hologram and its background, user's task, and the distance between the user and the hologram~\cite{kruijff2010perceptual}. 

\section{Conclusion}
\label{sec:Conclusion}
The ability to enjoy markerless AR through a smartphone is 
the result of decades of research and development in mobile hardware, computer graphics, and localization and mapping, among others. 
The recent expansion of AR capabilities to most phone models, along with no requirement for pre-printed markers, has brought AR experiences to a wider audience than ever before. However, much work remains for holograms to stay in place reliably, for stability to be robust to different environments and device movements, in order to support a greater variety of applications. The complex non-linearities of \mbox{VI-SLAM} combined with the specific challenges of smartphones (small field of view, movement patterns of handheld devices) mean that neither characterizing nor guaranteeing performance are solved problems. In this report we have shown that a good level of spatial stability of the holograms is achieved in the simplest cases (little device movement, and well-lit, textured environments), but that common, more challenging scenarios cause significant decreases in spatial stability,  
and improvements in the underlying technologies are required to address these issues. 



The scope of this work was single-session AR in static indoor environments. 
In 
our ongoing work, we are also considering other scenarios (e.g., outdoor spaces, specialized medical or industrial settings, or a busy classroom), in which extreme or variable lighting, textureless, dark or reflective surfaces, or dynamic objects may be more common. Inspired by initial work on measuring spatial consistency and latency in multi-user scenarios \cite{ran2020multi, he2019reducing}, we are currently adapting 
the methodology described in this report 
to measure the localization accuracy and latency of holograms designed to persist across multiple sessions using spatial anchors~\cite{AzureSpatialAnchors}. Performance in this case is highly dependent on visual conditions and their consistency over multiple days, weeks or months, and we will obtain quantitative insights into the environmental properties which best support persistent AR content. We have open-sourced our AR session measurement app HoloMeasure\footnotemark[1] to allow others to examine their own scenarios, and more rapidly reveal further insights on hologram stability in markerless AR.

\balance

\begin{acks}
Shreya Hurli, Priya Rathinavelu, and Alex Xu contributed to this work through undergraduate research projects. Experiments at the Health Innovation Lab at Duke University School of Nursing were conducted with the help of Ryan Shaw. This work was supported in part by the Lord Foundation of North Carolina and by NSF awards CSR 1903136, CNS 1908051, CAREER 2046072, and CAREER 1942700.
\end{acks}

\bibliographystyle{ACM-Reference-Format}
\bibliography{references}

\end{document}